\documentclass[twocolumn]{aastex63}

\hypersetup{linkcolor=red,citecolor=blue,filecolor=cyan,urlcolor=magenta}
\usepackage{newtxtext,newtxmath}
\usepackage[T1]{fontenc}
\usepackage{ae,aecompl}
\usepackage{graphicx} 
\usepackage{amsmath} 
\usepackage{amssymb} 
\submitjournal{ApJ}
\shorttitle{X-ray jets in hybrid morphology sources}
\shortauthors{Sebastian et al.}
\graphicspath{{./}{figures/}}
\begin{document}

\title{Investigating the origin of X-ray jets: A case study of four hybrid morphology MOJAVE blazars}
\correspondingauthor{Biny Sebastian}
\email{sebastib@purdue.edu}
\author[0000-0001-8428-6525]{Biny Sebastian}
\affiliation{Department of Physics and Astronomy, Purdue University, 525 Northwestern Avenue, West Lafayette, IN 47907, USA}
\author[0000-0003-3203-1613]{Preeti Kharb}
\affiliation{ National Centre for Radio Astrophysics - Tata Institute of Fundamental Research,
S. P. Pune University Campus, Post Bag 3, Ganeshkhind, Pune 411007, India}
\author[0000-0003-1315-3412]{Matthew L. Lister}
\affiliation{Department of Physics and Astronomy, Purdue University, 525 Northwestern Avenue, West Lafayette, IN 47907, USA}
\author[0000-0000-0000-0000]{Herman L. Marshall}
\affiliation{MIT Kavli Institute for Astrophysics and Space Research
Massachusetts Institute of Technology, 77 Massachusetts Avenue, Cambridge, MA 02139, USA}
\author[0000-0001-6421-054X]{Christopher P. O'Dea}
\affiliation{Department of Physics and Astronomy,
University of Manitoba, Winnipeg, MB R3T 2N2, Canada}
\author[0000-0000-0000-0000]{Stefi A. Baum}
\affiliation{Department of Physics and Astronomy,
University of Manitoba, Winnipeg, MB R3T 2N2, Canada}

\begin{abstract}
We have carried out {\it Chandra}, HST, and VLA observations of four MOJAVE blazars that have been previously classified as ``hybrid'' (FR\,I/II) blazars in terms of radio morphology but not total radio power. The motivation of this study is to determine the X-ray emission mechanism in jets, these being different in FR\,I and FR\,II jets. We detected X-ray jet emission with sufficient SNR in two blazars \textit{viz.} PKS\,0215+015 and TXS\,0730+504. We carried out spectral energy distribution (SED) modeling of the broad-band emission from the jet regions in these sources and found that a single synchrotron emission model is ruled out due to the deep upper limits obtained from HST optical and IR data. 
The IC-CMB model can reproduce the X-ray jet emission in both sources although the model requires extreme jet parameters. 
Both our sources possess FR\,II like radio powers and our results are consistent with previous studies suggesting that radio power is more important than FR morphology in determining the emission mechanism of X-ray jets. 
\end{abstract}



\section{Introduction} \label{sec:intro}

The \textit{Chandra} X-ray Observatory has made it possible to detect X-ray emission from spatially resolved jets of AGN, with X-ray jets being discovered in abundance in radio-loud AGN \citep{Sambruna2004,Marshall2005}. Several studies revealed systematic trends in the X-ray jets based on their Fanaroff-Riley (FR) classification 
\citep{Fanaroff1974}.
While the X-ray emission from the majority of FR\,I jets can be explained using the synchrotron mechanism, several FR\,II jets reveal a convex spectral energy distribution (SED) which requires additional mechanisms to explain their high X-ray flux levels, for example, the inverse-Compton mechanism, where the seed photons came from the cosmic microwave background \citep[IC/CMB;][]{Harris2006}.


The Fanaroff-Riley classification was originally based on the location of the brightest feature on either side of the radio core \citep{Fanaroff1974}. In FR\,I or edge-darkened sources, the jets flare into plumes close to the core and the surface brightness decreases from there towards the edge of the sources. On the other hand, the jets in FR\,II or edge-brightened sources remain stable and collimated till they terminate in the bright features called ``hostspots''.
It was also noted that the radio power at 178~MHz, was systematically different for the two classes, with FR\,IIs having greater radio luminosities than FR\,I sources (see \cite{Bridle1984} for additional discussion).
More recently, \cite{Mingo2019} showed using a larger sample that the dividing power is not as rigid as was previously assumed. They find that majority of the sources possessing 150~MHz radio luminosities (L$_{150}$) greater than $2\times10^{25}$~W~Hz$^{-1}$ are FR\,II type sources. 
The current paradigm is that the jets in the FR\,I sources decelerate on the kpc scales \citep{Parma1994,Laing2002} whereas the  jets in the FR\,II radio sources remain relativistic out to the terminal hot spots \citep{Hardcastle2008}. 
Explanations for the FR dichotomy have included differences in the central black hole mass, spin, accretion rates/modes, jet composition, and the external environment \citep[e.g.,][]{Celotti1993,Baum1995,Meier1999,Ghisellini2001}. 

A tiny fraction of radio-powerful AGN \citep[$<$1\%;][]{Gopal-Krishna2000,Gawronski2006} however, exhibit an FR\,I-like jet on one side of the radio core and an FR\,II-like jet with a terminal hotspot, on the other. These sources are referred to as ``hybrid'' FR sources. As intrinsic differences in the central engine (e.g., black hole mass, spin, accretion rate, jet composition) can be ruled out for hybrid sources, they could provide {the key to finally understanding the FR dichotomy.} \citet{Gopal-Krishna2000} showed that the total radio luminosities of these sources were close to the dividing line between FR\,Is and FR\,IIs. Asymmetries in the surrounding medium, which could preferentially decelerate and de-collimate one jet, were proposed as an explanation for the hybrid morphology sources. 
In a recent study by \cite{Harwood2020}, it was concluded that the hybrid morphology sources are intrinsically FR\,II sources which appear FR\,I on one side due to a combination of projection effects and outer lobes not being aligned parallel to the inner jet. 

For the handful of hybrid FR sources observed with \textit{Chandra} and HST, systematic trends in the X-ray emission mechanisms have, however, not been observed. \citet{Kharb2015} and \citet{Stanley2015} presented the results from {Chandra} and {HST} imaging of three MOJAVE \citep[Monitoring Of Jets in AGN with Very Long Baseline Array (VLBA) Experiments;][]{Lister2009} hybrid blazars, along with multi-waveband data of ten hybrid sources from the literature.

It was found that the majority of the hybrid sources had ``high'' radio powers. While the synchrotron emission mechanism was enough to explain the X-ray jet emission in low-power sources the IC-CMB model was invoked to explain the X-ray emission in hybrid sources when the radio power was high \citep[see figure~2 in ][]{Kharb2015}. We use the dividing line of 5$\times$10$^{26}$~W~Hz$^{-1}$ at 150~MHz to define ``high'' radio powers. It is also possible that these high-power hybrid morphology
sources are projected FR\,II sources as suggested by \cite{Harwood2020}, which could explain both their high powers as well as their convex SEDs.


Although IC-CMB model is widely favored to explain the convex SED in FR\,II type sources \citep{Sambruna2002,Sambruna2004,Miller2006,Tavecchio2007,Marshall2011,Perlman2011,Kharb2012,Stanley2015} a few authors have pointed out various shortcomings of the model. For instance, the requirement of highly relativistic jets on kpc-scales to reproduce the X-ray flux levels and the observed spatial offsets between the X-ray jet knots from the radio features raise concerns about this simple picture  \citep{Hardcastle2006}. While it is true that most X-ray jets in FR\,II sources follow the IC-CMB emission mechanism, there are exceptions. For example, \cite{Hardcastle2006} find that the jet emission in Pictor\,A, despite being an FR\,II source, is consistent with synchrotron mechanism rather than IC-CMB. More recently, the IC-CMB model has been ruled out in several sources because the minimum-detected $\gamma$-ray fluxes obtained from \textit{Fermi} telescope surveys are lower than that predicted by the model \citep{Meyer2014,Meyer2015,Meyer2017,Breiding2017}. 
Additionally, \cite{Marshall2018} found that the X-ray jets in quasars did not follow the trend expected with redshift according to the IC-CMB model. However, recently \cite{Meyer2019} have reported that the low-state gamma-ray flux density of two sources, a low-power BL Lac, and a powerful quasar  are consistent with the IC-CMB mechanism.

We present the broadband SED modeling study of a sample of four hybrid morphology MOJAVE blazars which were observed with \textit{Chandra} and {HST}. The sources were classified based on A-array Very Large Array (VLA) images at 1.5~GHz. Those sources which showed evidence for a hotspot on one side and none on the other side were classified as hybrid morphology sources. These are the four sources out of the ten hybrid sources identified in the complete radio-selected MOJAVE sample which did not have deep \textit{Chandra} observations (see Table~\ref{tab1} for more details). The sources which did not show jet detection in the pilot {\textit{Chandra}} study \citep[4C~+67.05; ][]{Hogan2011}, or had a jet that was smaller than the \textit{Chandra} point spread function (4C~+39.25) were excluded. 
We point out some subtleties underlying our classification strategy here. The hotspot of one of our sample sources gets resolved out in the images with resolution higher than that used for the initial classification. We used the presence of a hotspot rather than the edge brightening as the criteria to classify sources as FR\,II type since most FR\,II sources show unresolved compact components in high resolution images as well \citep{ODea2009,Fernini2014}. Some of our sources show edge-brightening on the FR\,I side even though a hotspot is absent. These lobes resemble the ``fat-doubles'' in large radio galaxy samples \citep{Owen1989}. “Fat doubles” may show bright outer rims of radio emission (resembling edge-brightened sources) but have roundish diffuse lobes; their optical properties resemble those of FR\,Is rather than FR\,IIs.
The aim of this study is to gain some insight into the preferred emission models and the causes for the trends based on the FR dichotomy. 

In section~\ref{sec:obs}, we describe our X-ray, optical and radio observations and the data analysis procedures that were used. We present the results including the description of each source and the details about the SED modeling in Section~\ref{sec:results}. We discuss the implications of the SED modeling results, the viability of the IC-CMB model, and the alternative mechanisms of emission in Section~\ref{sec:discussion}. We present our conclusions and discuss the directions for future work in Section~\ref{sec:conclusion}.

Throughout this paper, we use $H_0$=73~km~s$^{-1}$Mpc$^{-1}$, $\Omega_{m}=0.27$ and $\Omega_{vac}=0.73$.
\section{Observations and Data Analysis}
\label{sec:obs}
\begin{table*}
\begin{center}
\caption{Our sample of sources}
\begin{tabular}{cccccc} \hline \hline
%
Source & R.A.&Dec. &Redshift& $\beta_\mathrm{app}$ &VLBA P.A.\\
& && & &range($\degr$) \\
\hline
PKS\,0215+015   & 02:17:48.95   &$+$01:44:49.7    & 1.715 & 25.3$\pm$1.2$^a$ & 99.9$-$105.9 \\
TXS\,0730+504   & 07:33:52.5205 &$+$50:22:09.062  & 0.72 & 16.09$\pm$0.53$^a$ & 209.6$-$212.3\\
PKS\,1036+054   & 10:38:46.7798 &$+$05:12:29.085  & 0.473 & 6.3$\pm$1.3$^a$ & 352.1$-$354.9\\
PKS\,1730$-$130 & 17:33:02.7057 &$-$13:04:49.547  & 0.902 &27.3$\pm$1.0$^b$ &352.4$-$26.5 \\
\hline
\label{tab1}
\end{tabular}
\\{Note: $\beta_\mathrm{app}$ is the apparent jet speed in units of the speed of light and corresponds to the maximum jet speed observed in these sources} $^a$-\cite{Lister2019}; $^b$-\cite{Lister2021}
\end{center}
\end{table*}


\begin{table*}
\begin{center}
\caption{Observation log and observed parameters.}
 \begin{tabular}{cccccccc}
\hline \hline
{\bf VLA radio observations} \\ \hline
Object & VLA Array & Observation & $\nu_{\rm cen.}$ & Beam, PA & Image peak flux & Image r.m.s  & On source\\
 & Configuration & date & (GHz) & (arcsec, $\degr$) & density (Jy) & (mJy~beam$^{-1}$)& time (min) \\
\hline

PKS\,0215+015 & A &	 2012 Dec 31 & 6.0	 &   0.34$\times$0.25, 32.68 & 0.96 & 2.6$\times$10$^{-1}$ & 6.87	\\
         & A &	 2012 Dec 31 & 1.5	 &    1.26$\times$0.96, 20.93 & 1.02	 &	8.3$\times$10$^{-3}$ & 6.33	\\
         & B &	 2013 Dec 18 & 5.5	 &   1.44$\times$0.93, 35.99 & 0.65 & 1.33$\times$10$^{-1}$ & 1.4	\\
         & B &	 2015 Feb 20 & 1.5	 &    3.51$\times$2.86, -13.2 & 0.73	 &	2.04$\times$10$^{-1}$ &	 11.1\\
         & B &   2019 Jun 16 &  10.0 &  	0.9$\times$0.7,	-14.66 &	1.8      & 9.00$\times$10$^{-1}$ & 4.5	\\
TXS\,0730+504 & A &	 2007 Aug 13 & 5.5	  & 0.32$\times$0.29, -52.83 & 0.42 &	1.24$\times$10$^{-1}$ &	2.0\\
         & A &	 2015 Aug 28 & 1.42 &    1.58$\times$1.07, -44.26 & 0.61 &	3.4$\times$10$^{-1}$ &	21.0\\
         & B &	 2015 Mar 21 & 1.52 &   3.37$\times$2.84, -167.99 & 0.45 &	2.14$\times$10$^{-1}$ & 3.9	\\
PKS\,1036+054 & A &	 2015 Jul 16 & 5.5	 &   0.44$\times$0.28, 45.64 & 1.23	 &	1.8$\times$10$^{-1}$ &	1.05\\
         & A &	 2015 Jul 16 & 1.52 &    1.47$\times$1.01, 45.69 & 0.91	 &	2.48$\times$10$^{-1}$ &	2.95\\
         & B &	 2015 Mar 21 & 5.5	 &   1.66$\times$0.95, -47.05 & 1.41 &	3.43$\times$10$^{-1}$ &	1.05\\
         & B &	 2015 Mar 21 & 1.52 &    5.21$\times$3.0, -46.16 & 1.01 &	2.06$\times$10$^{-1}$ &	2.8\\


\\
\end{tabular}
%


\begin{tabular}{cccccc} \hline \hline
\\
{\bf \textit{Chandra} X-ray observations} \\ \hline
Source & Observation & OBSID & Exposure & Net Count rate & \\
& Date& &(s) & ks$^{-1}$ \\
\hline
PKS\,0215+015   & 2019 Nov 03 & 22564 & 22735.8 & 102.6 \\
                & 2019 Nov 04 & 23063 & 20118.6 & 100.4 \\
TXS\,0730+504   & 2020 Jan 29 & 22565 & 13667.2 & 79.9 \\
                & 2020 Jan 29 & 23134 & 28178.3 & 77.7 \\
PKS\,1036+054   & 2020 Feb 07 & 22566 & 22736.7 & 54.0 \\
                & 2020 Feb 08 & 23144 & 22546.7 & 51.8 \\
PKS\,1730$-$130 & 2020 Jun 16 & 22567 & 27330.6 & 147.4 \\
                & 2020 Jun 18 & 23284 & 16387.4 & 156.6 \\

\end{tabular}
\\
\begin{tabular}{ccccccc} \hline \hline
{\bf \textit{Hubble} optical observations} \\ \hline
Source & Observation & Instrument & Aperture & Filter & Pivot & Exposure time\\
& Date& & & & Wavelength (\AA) & time(s)\\
\hline

PKS\,0215+015   & 2020 Jul 06 & WFC3 & IR-UVIS-FIX & F160W & 15369.1 & 2408.8 \\
                & 2020 Jul 06 & WFC3 & UVIS-IR-FIX & F475W & 4772.6 & 2486.0 \\
TXS\,0730+504   & 2020 May 12 & WFC3 & IR-UVIS-FIX & F160W & 15369.1 & 2708.8 \\
                & 2020 May 12 & WFC3 & UVIS-IR-FIX & F475W & 4772.6 & 2767.0 \\
PKS\,1036+054   & 2020 May 14 & WFC3 & IR-UVIS-FIX & F160W & 15369.1 & 2408.8 \\
                & 2020 May 14 & WFC3 & UVIS-IR-FIX & F475W & 4772.6 & 2579.0 \\
PKS\,1730$-$130 & 2020 May 14 & WFC3 & IR-UVIS-FIX & F160W & 15369.1 & 2408.8 \\
                & 2020 May 14 & WFC3 & UVIS-IR-FIX & F475W & 4772.6 & 2583.0 \\


\hline
\label{tab:obs}
\end{tabular}
\end{center}
\end{table*}

\subsection{\textit{Chandra} Data}
\label{sec:chandra_dataanal}

The four sources in our sample were observed using the Advanced CCD Imaging Spectrometer Back Illuminated chip (ACIS-S3) for optimized low energy response. To reduce the effects of pileup from the bright AGN cores, saturation, and background, we employed the very faint (VFAINT) telemetry and the 1/8 subarray mode. The spacecraft roll angles were constrained to prevent the kpc-scale jet from coinciding with the trail of charge transfer from the core. The {\it Chandra} data were acquired in two sub-exposures. The details of the observations are summarized in Table~\ref{tab:obs}.

\begin{figure*}
 \centering
 \includegraphics[height=5.9cm,trim = 0 0 0 0 ]{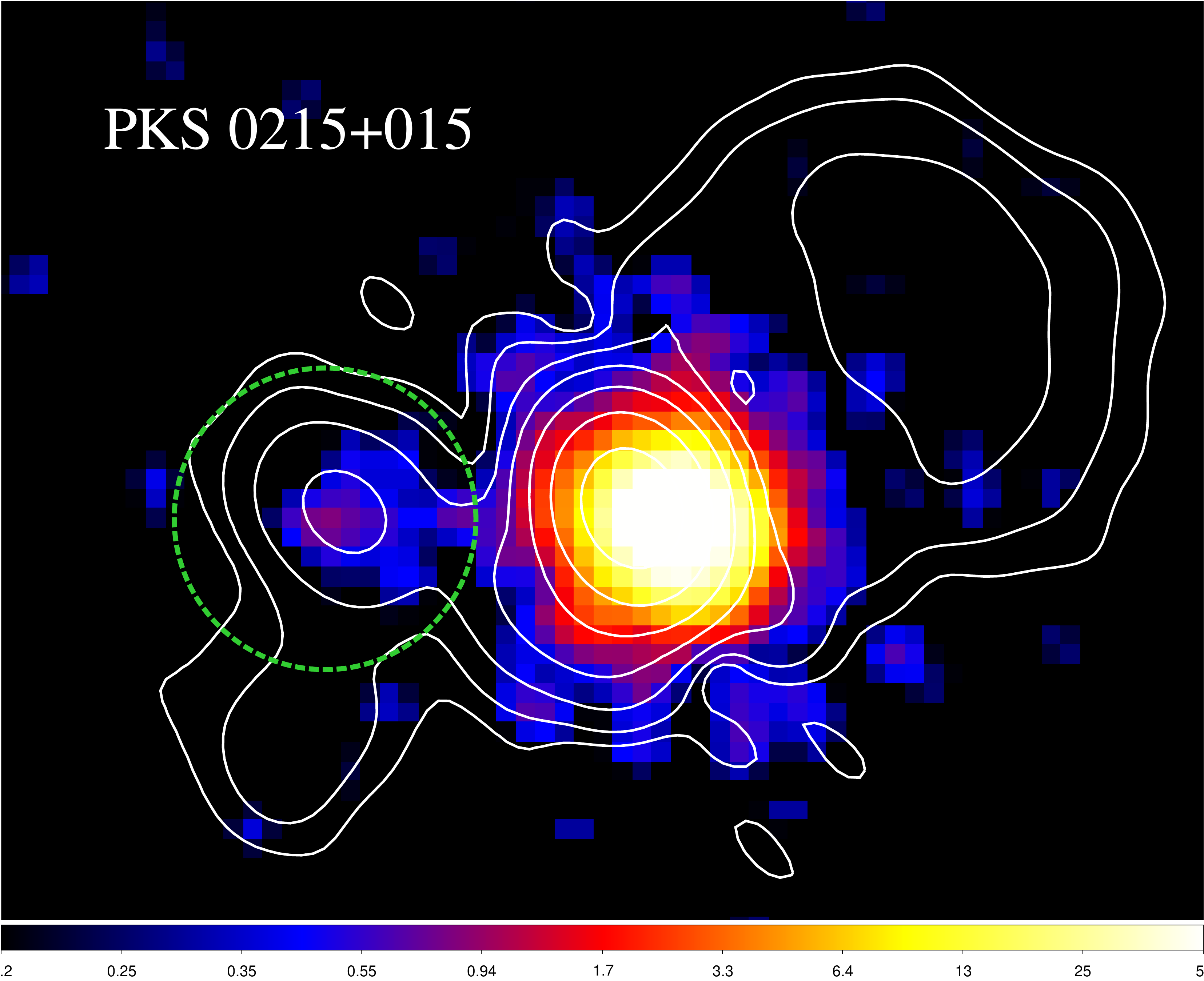}
 \includegraphics[height=5.9cm,trim = 0 0 0 0 ]{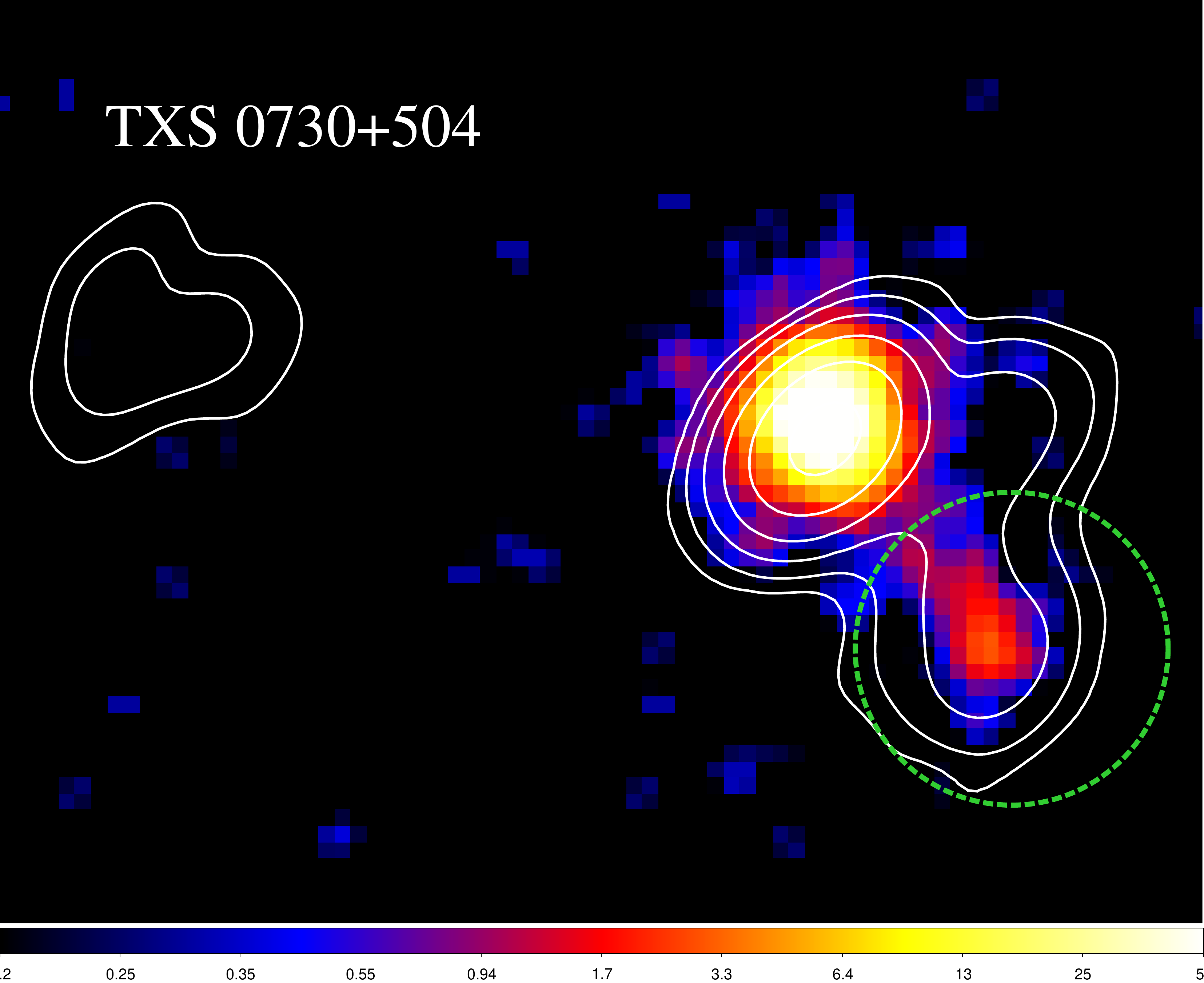}
 \\\includegraphics[height=5.9cm,trim = 0 0 0 0 ]{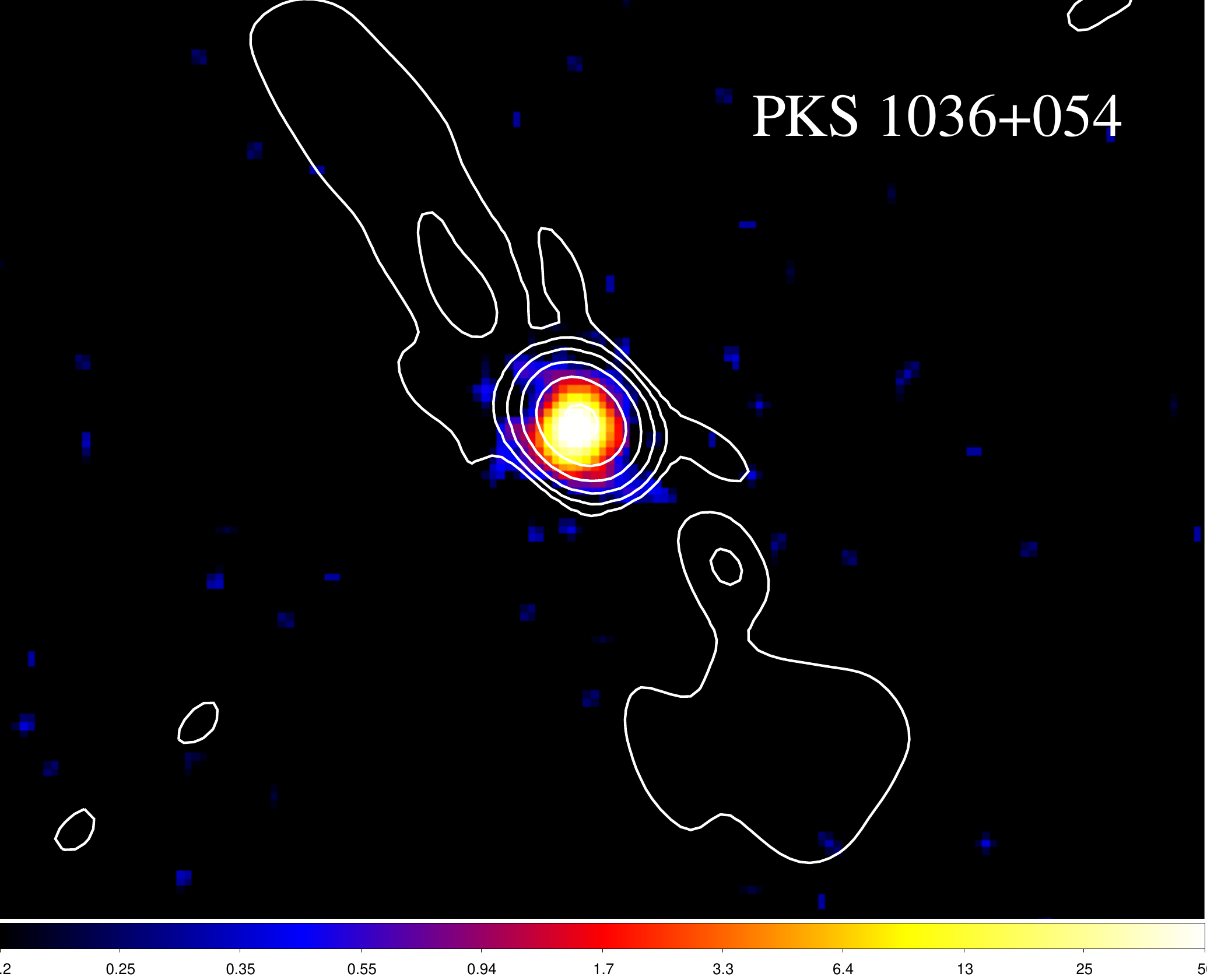}
 \includegraphics[height=5.9cm,trim = 0 0 0 0 ]{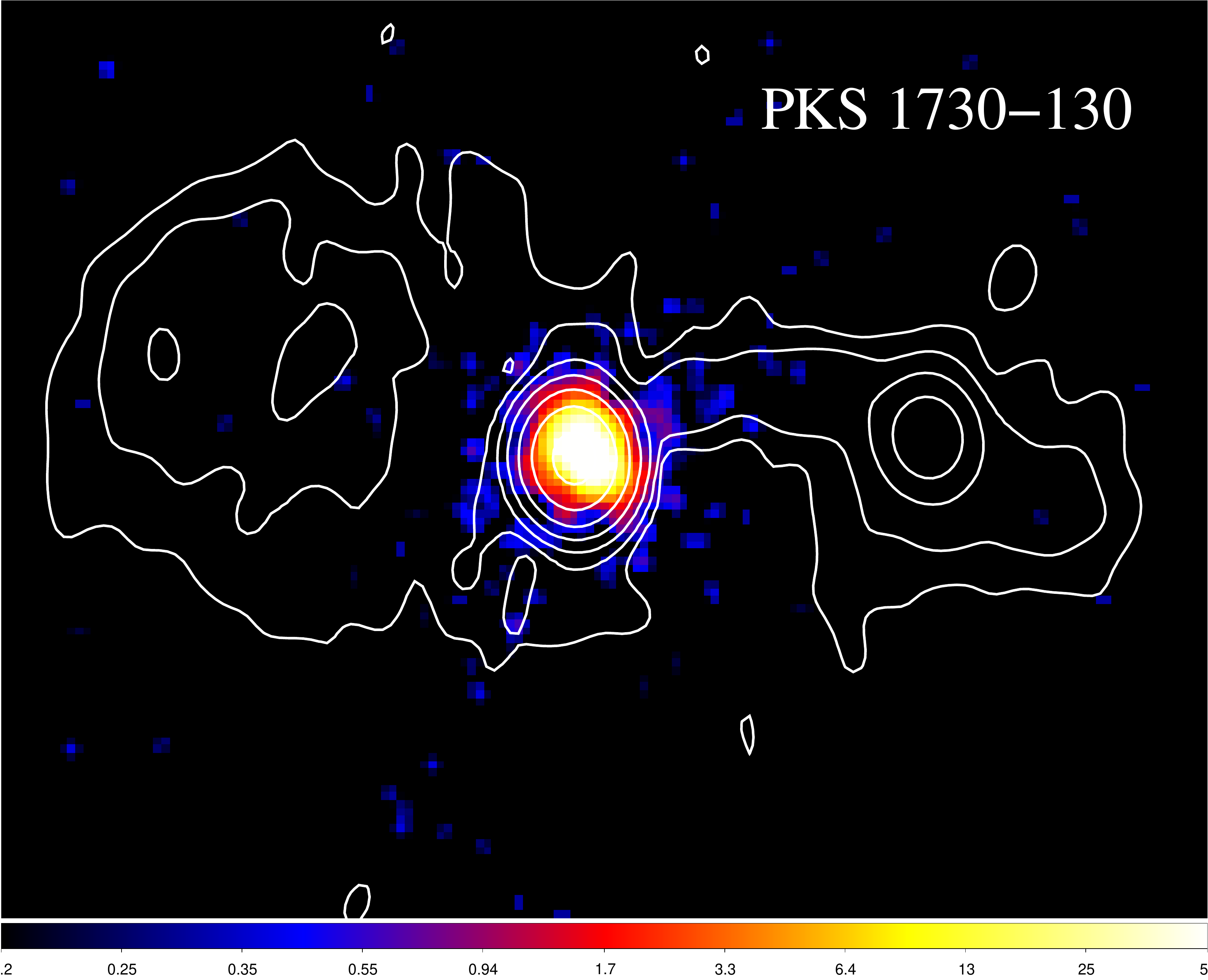}
 \caption{The \textit{Chandra} X-ray color images of PKS\,0215+015 (top left), TXS\,0730+504 (top right), PKS\,1036+054 (bottom left), and PKS\,1730$-$130 (bottom right) with total intensity contours of the VLA A-array 1.4~GHz data superposed on top. The X-ray images were binned to 0.5 times the original pixel size and smoothed with a Gaussian kernel of radius 2. The green circle represents the jet region chosen for SED modeling. The lowest contour level is given by 3$\sigma$, where $\sigma$ is the rms noise of the image and the following levels are given by a factor of four times the previous level (see Table~\ref{tab:obs}.) The radio total intensity image of PKS\,1730$-$130 is presented in \cite{Kharb2010}.}
 \label{fig:x-ray}
\end{figure*}

We reprocessed the data from the two sub-exposures separately using the {\tt chandra\_repro} script in \textit{Chandra} Interactive Analysis of Observations ({\tt CIAO}) package version 4.12.1 with calibration
database version (caldb) 4.9.3 and combined them using the {\tt merge$\_$obs} script. 
We used the SAOImage DS9 tool to generate the \textit{Chandra} X-ray and Very Large Array (VLA) radio overlays. We smoothed the images using a Gaussian kernel radius of 2 and adjusted the scale-limits to 0.2 - 50. Figure~\ref{fig:x-ray} shows \textit{Chandra} color images of all the sources overlaid with VLA contours.

Before the X-ray spectral analysis, we filtered the data to limit the energy range from 0.5 - 10 keV. We generated circular region files for the core, the jet, and the background from a source free location using the DS9 tool. We then made use of the {\tt CIAO} script {\tt specextract} to generate the spectrum for the jet and the core regions for the two sub-exposures separately.
The spectral fitting for the cores of all the sources and the jet regions in the case of PKS\,0215+015 and TXS\,0730+504 were carried out with the {\tt XSPEC} package (HEASOFT version 12.11.1). The {\tt XSPEC} model phabs(powerlaw) which represents absorbed power law provides reasonable fits to the data. From the power law parameters, namely the photon index and the normalization factor, the unabsorbed flux of the jet region was calculated at three frequencies (0.5, 3.7, and 7.0 keV) for the two sources which show extended jet features. These data points were used for the broadband SED modeling described in Section~\ref{Sec:sedmodel}.

We also studied the diffuse X-ray emission around the hybrid-morphology sources to detect any asymmetries in the hot gas environment that surrounds the two different FR lobes. We defined two 120$\degr$ wide annulus sector regions centered on the jet axes using the PANDA shape in DS9. The inner radius of the annulus was chosen to be the outer edge of the lobe as seen in the radio images to avoid contamination from the jet. The outer radius was chosen to be 1.5 times the inner radius. We estimated the difference in the total counts between the FR\,I and FR\,II side regions. We also estimated the mean and standard deviation in the difference from random sampling across the X-ray image. The differences between the two regions in PKS\,0215+015, TXS\,0730+504 and PKS\,1036+054 were not statistically significant. On the other hand, the difference between the west and east lobe environments in PKS\,1730$-$130 was $\sim$2.5$\sigma$, where $\sigma$ is the standard deviation of the differences. Using a similar approach we also estimated the difference in the net counts within the inner sector regions for PKS\,1730$-$130, which exhibited a similar trend as the outer environment.

\subsection{\textit{HST} Data}

All the four sources were observed using the Wide Field Camera 3 (WFC3) on board HST with two wideband filters, {\it viz.} F160W and F475W filters for one orbit each. More details of the observations are provided in Table~\ref{tab:obs}.
We downloaded all the images from the Space Telescope Science Institute archive. These images were already calibrated using the AstroDrizzle processing, which is the python realization of MultiDrizzle \citep{Koekemoer2003}. 

\begin{figure*}
 \centering
 \includegraphics[height=6.6cm,trim = 0 0 0 0 ]{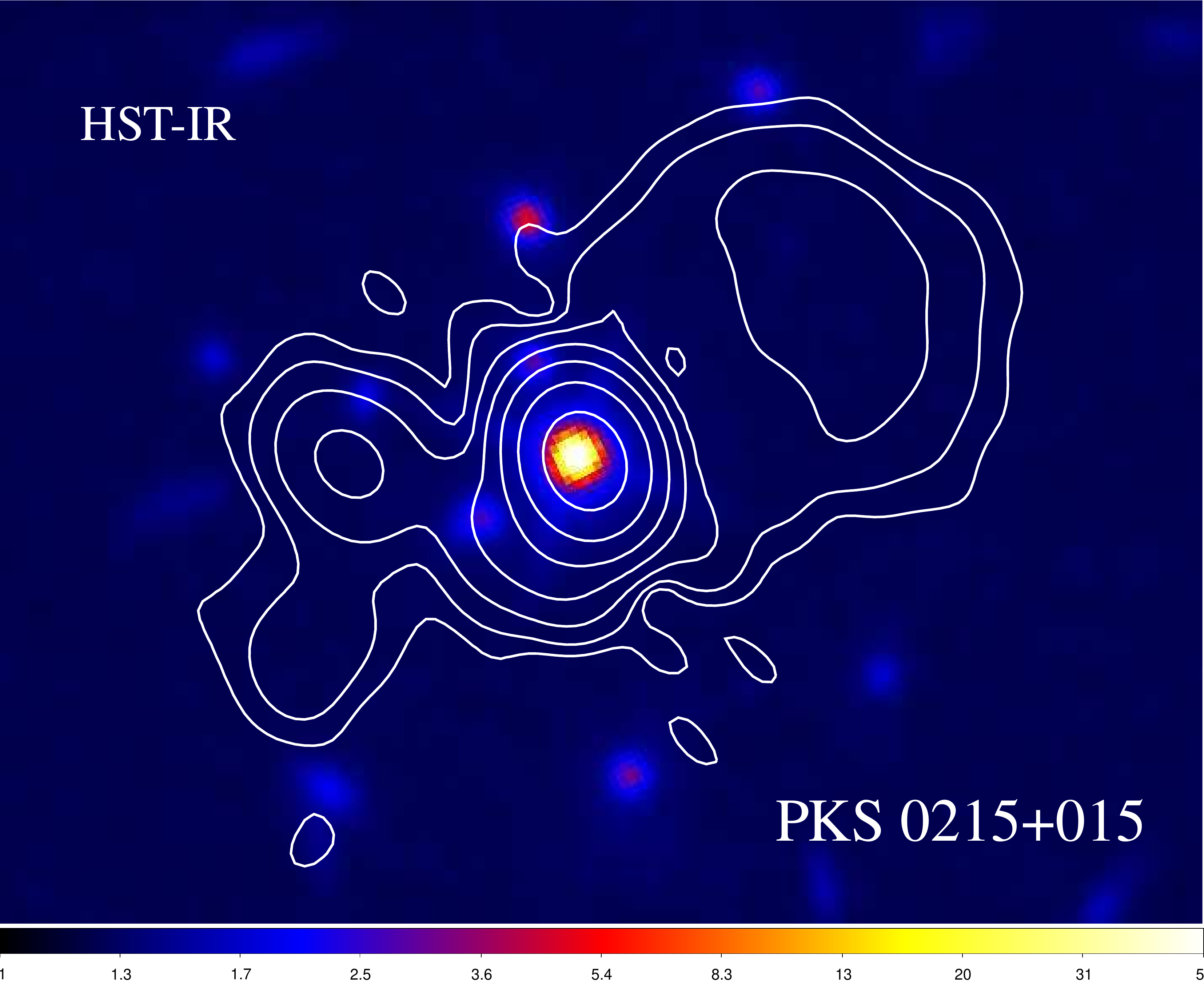}
 \includegraphics[height=6.6cm,trim = 0 0 0 0 ]{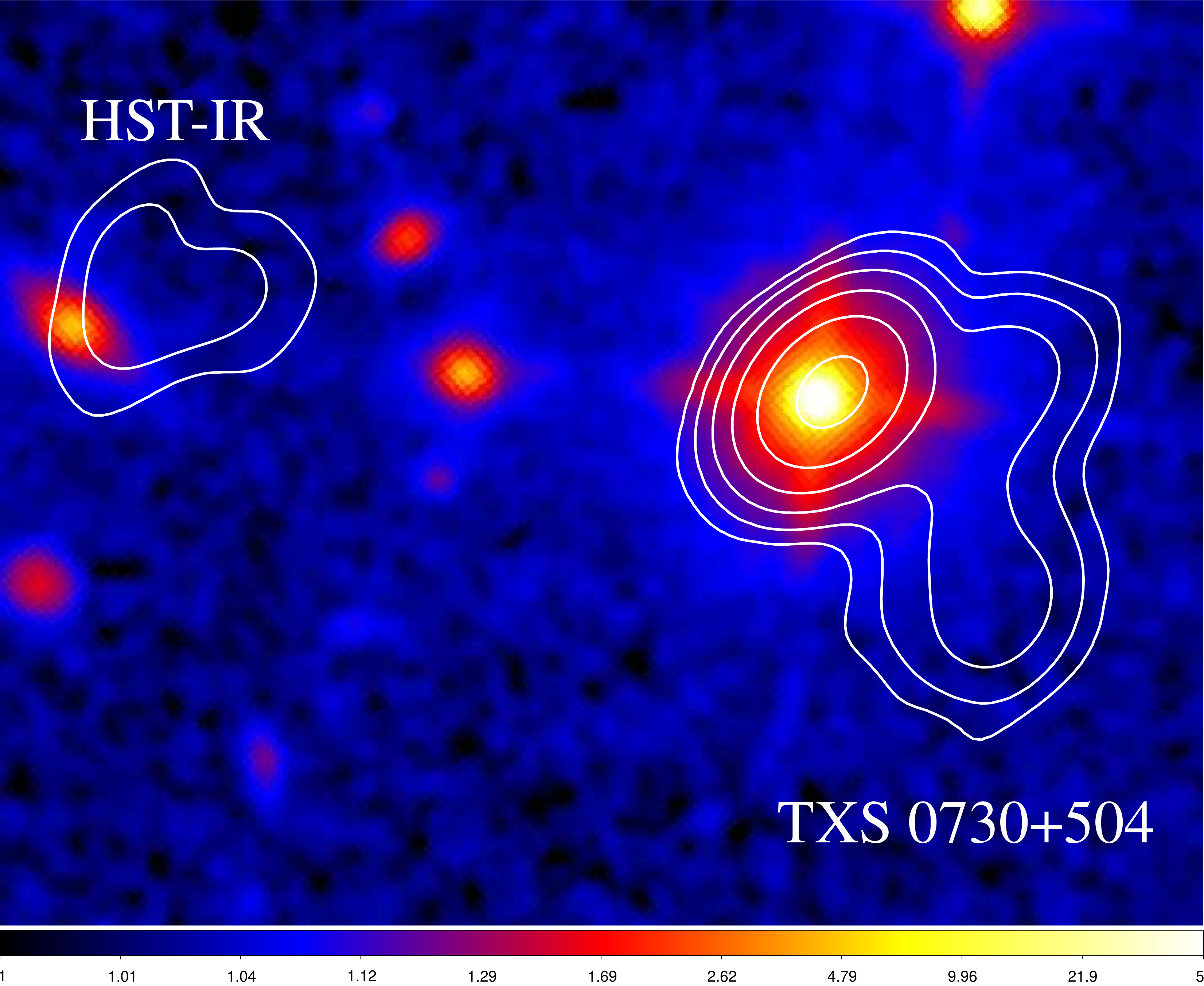}
 \includegraphics[height=6.6cm,trim = 0 0 0 0 ]{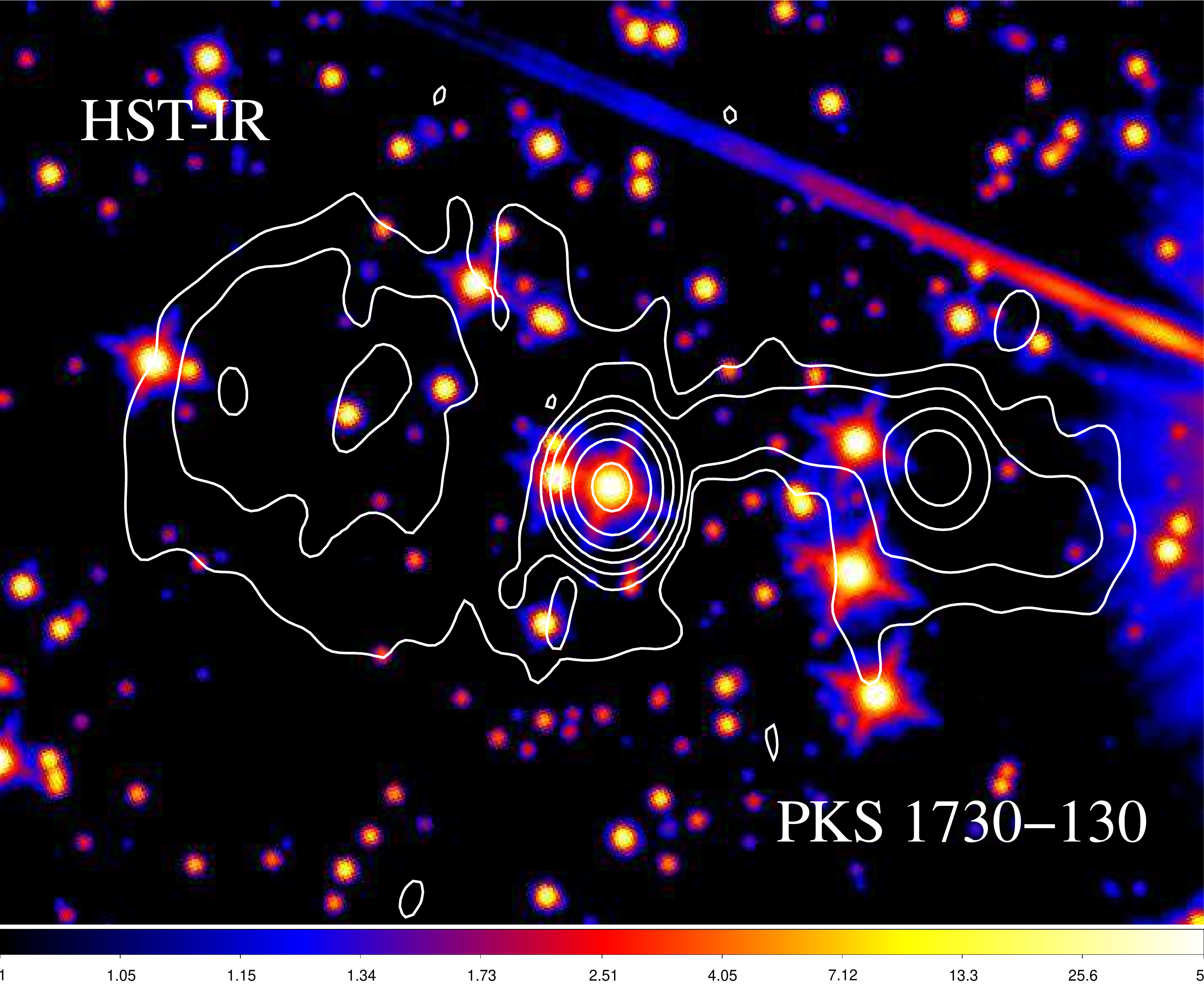}
 \caption{HST infrared color images of PKS\,0215+015 (top left), TXS\,0730+504 (top right), and PKS\,1730$-$130 (bottom) with total intensity contours of the VLA A-array 1.4~GHz data superposed on top. The radio contours are the same as in figure~\ref{fig:x-ray}. The lowest contour level is given by 3$\sigma$, where $\sigma$ is the rms noise of the image and the subsequent levels are four times the previous level (see Table~\ref{tab:obs}.) The radio total intensity image of PKS\,1730$-$130 is presented in \cite{Kharb2010}.}
 \label{fig:IR}
\end{figure*}

\begin{figure*}
 \centering
 \includegraphics[height=5.9cm,trim = 0 0 0 0 ]{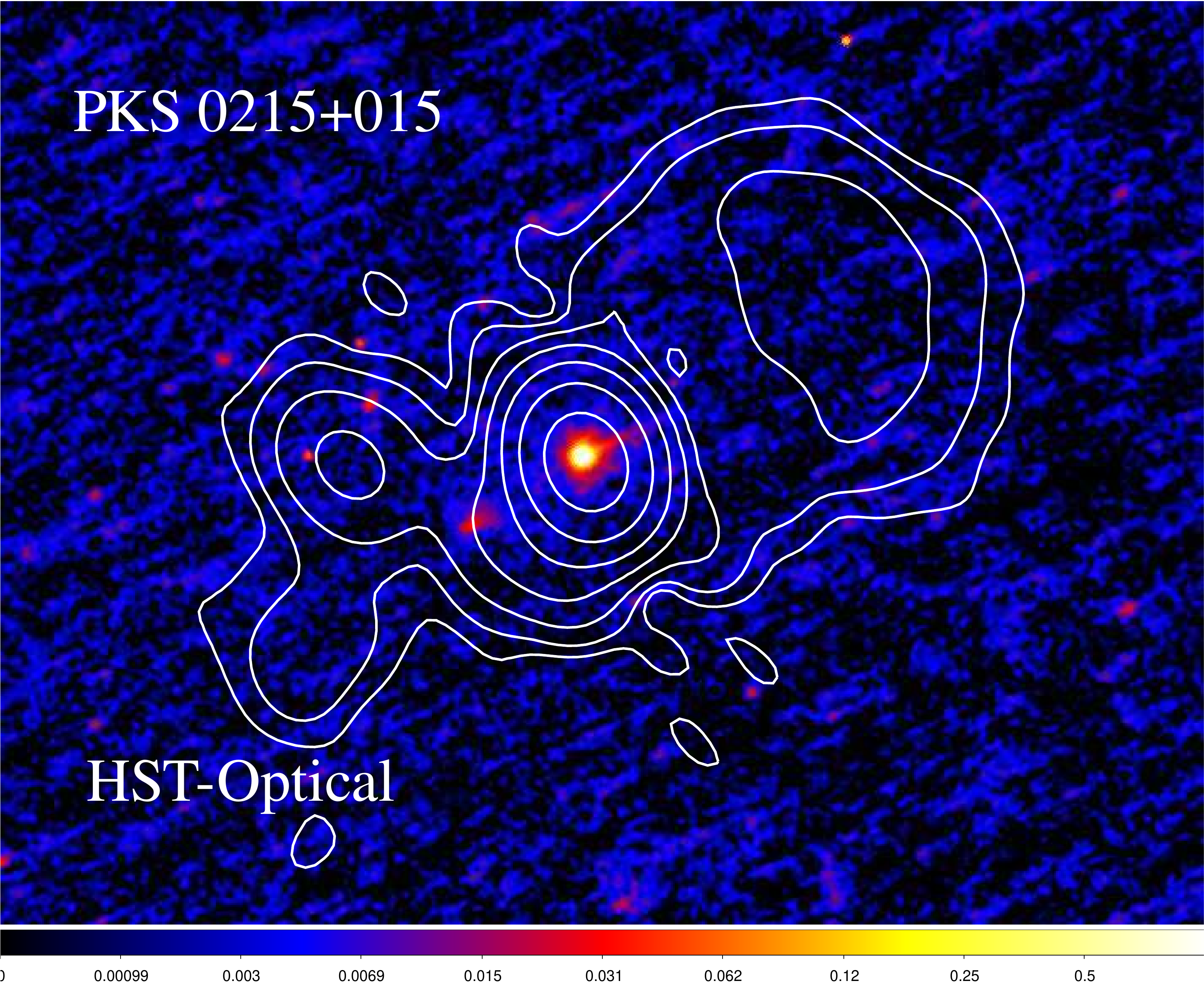}
 \includegraphics[height=5.9cm,trim = 0 0 0 0 ]{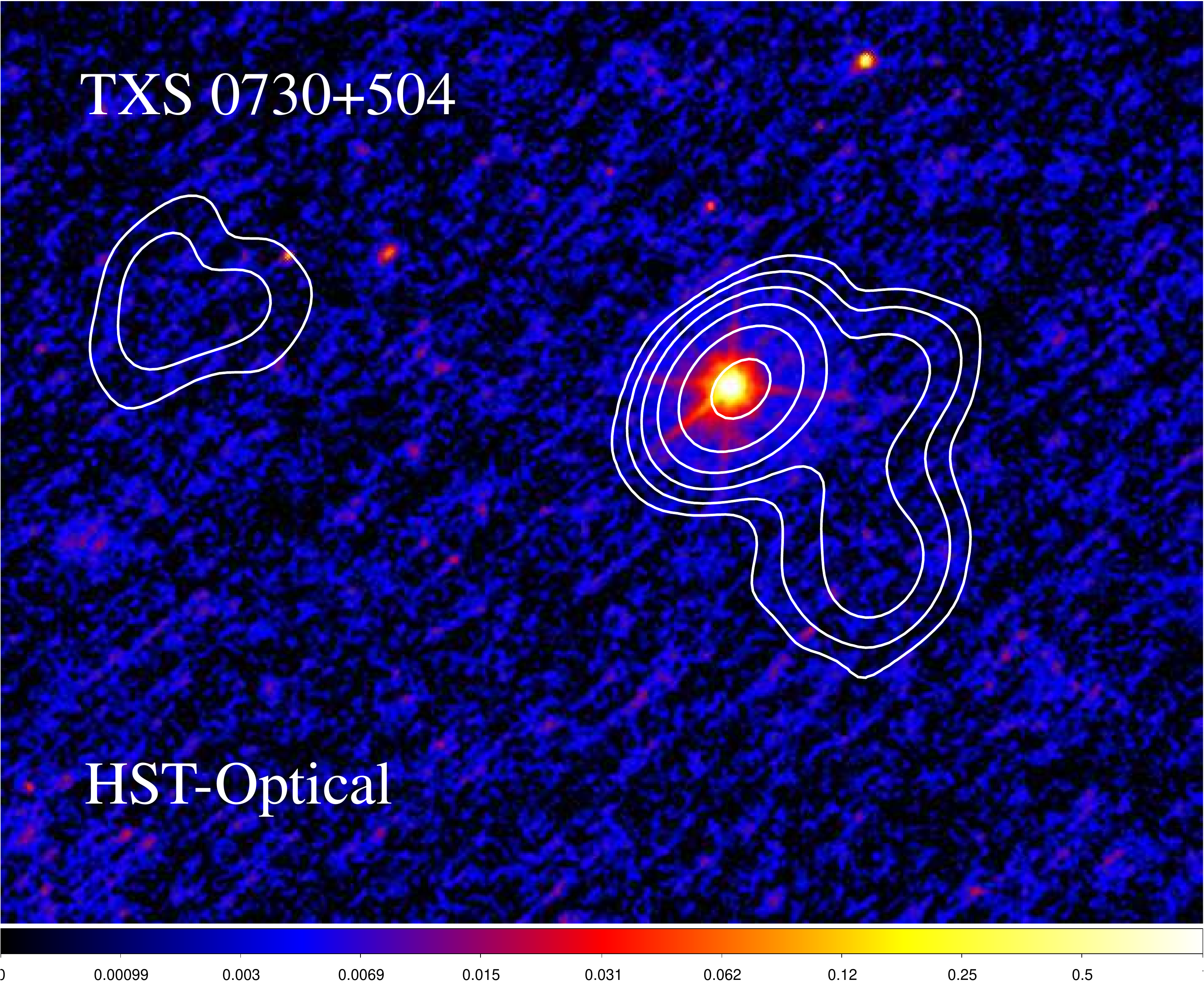}
 \\\includegraphics[height=5.9cm,trim = 00 0 0 0 ]{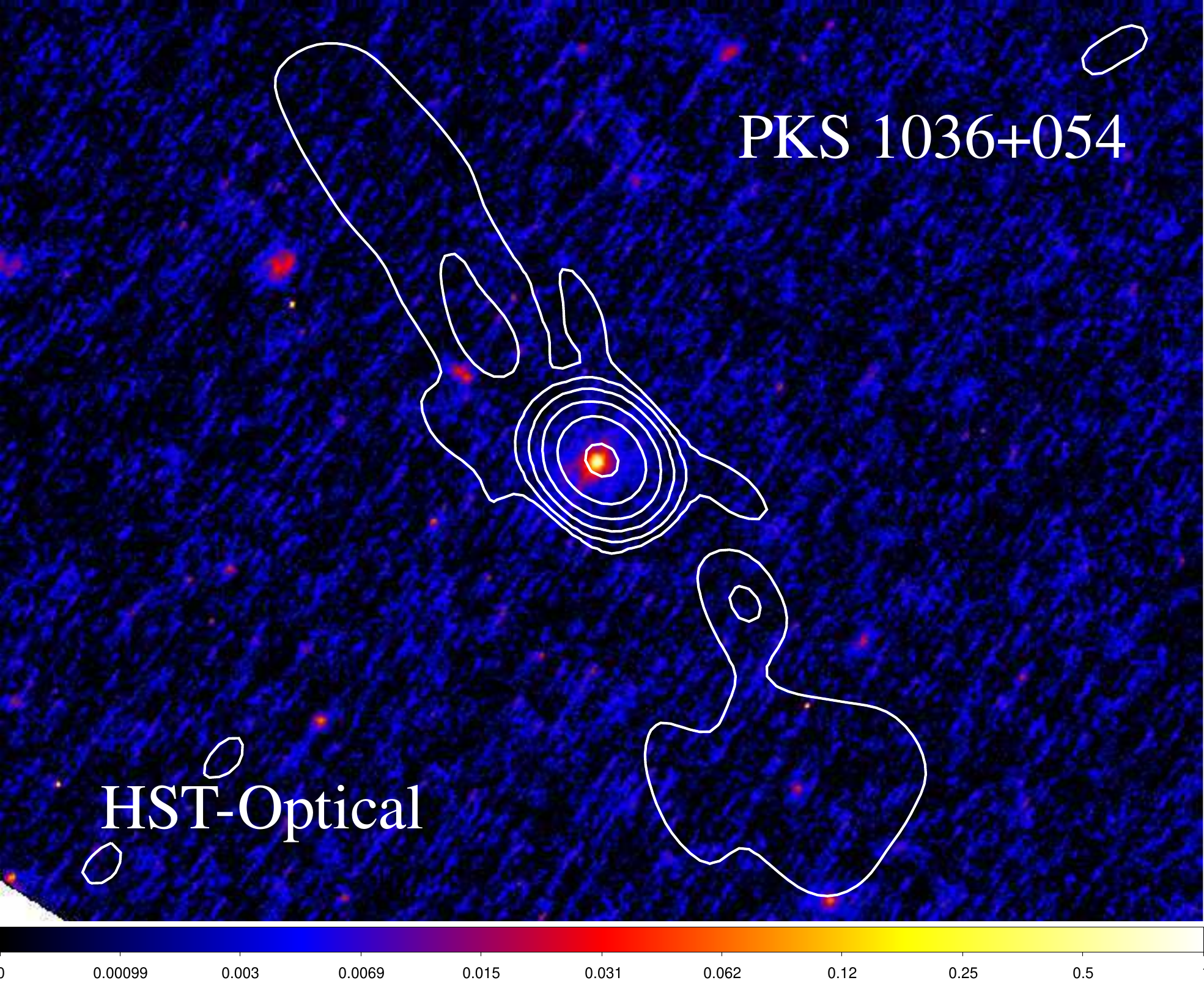}
 \includegraphics[height=5.9cm,trim = 0 0 0 0 ]{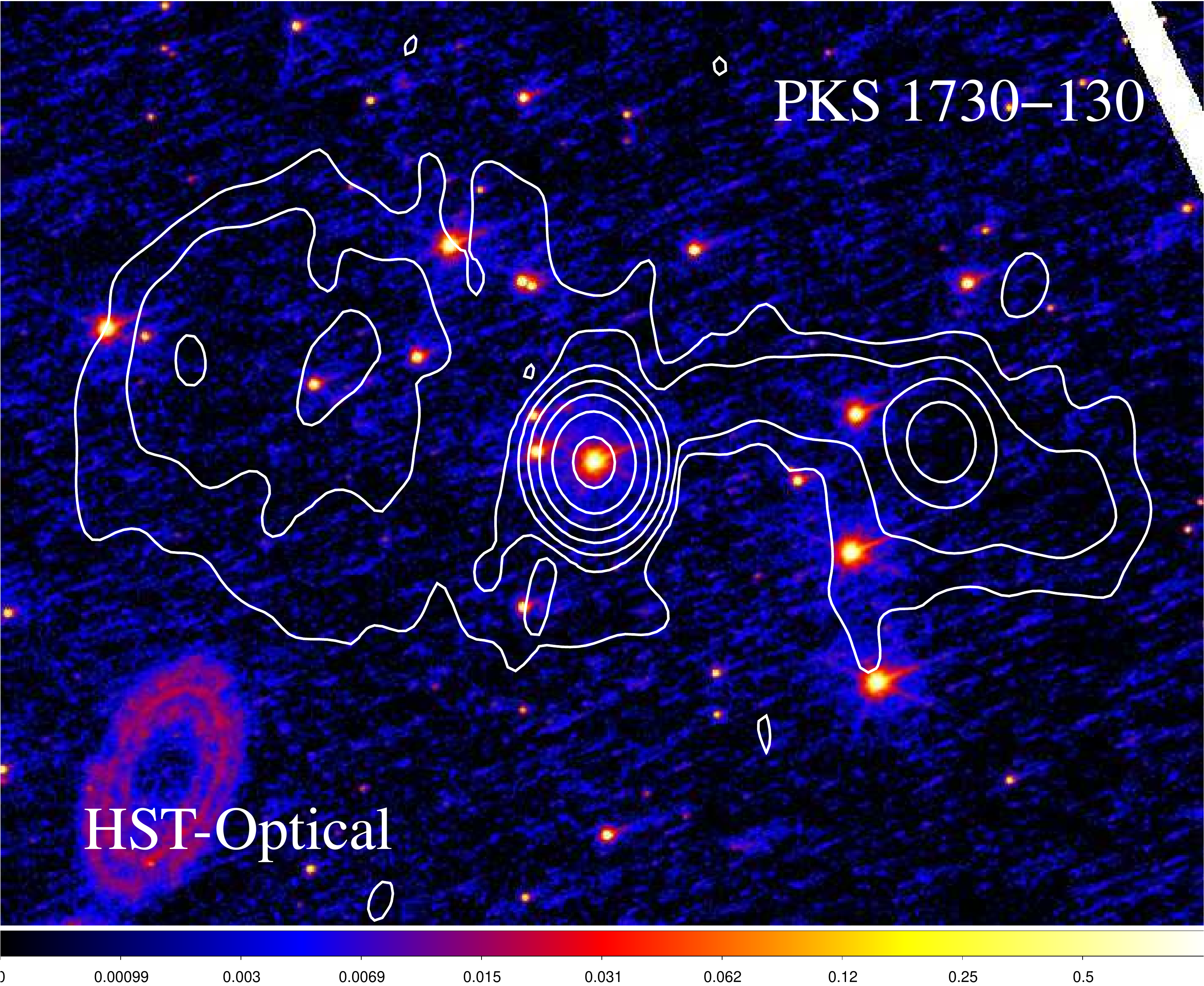}
 \caption{HST optical color images  of PKS\,0215+015 (top left), TXS\,0730+504 (top right), PKS\,1036+054 (bottom left), and PKS\,1730$-$130 (bottom right) with total intensity contours of the VLA A-array 1.4~GHz data superposed on top. The radio contours are the same as in figure~\ref{fig:x-ray}. The lowest contour level is given by 3$\sigma$, where $\sigma$ is the rms noise of the image and the subsequent levels are four times the previous level (see Table~\ref{tab:obs}.) The radio total intensity image of PKS\,1730$-$130 is presented in \cite{Kharb2010}.}
 \label{fig:optical}
\end{figure*}

We do not detect optical or infrared emission at the locations of the jet or lobes in any of the sources. In PKS\,0215+015 and TXS\,0730+504, where we have X-ray detections from jet regions, it is important to estimate the upper limit from the optical waveband to differentiate between the various emission models. To estimate the upper limit on the flux density values from the jet regions, we chose the same regions used for estimating the X-ray and radio flux densities. We used DS9 to estimate the rms value of the counts per pixel from the selected region. We then converted the units to the counts per beam by using the scaling suggested in the Wide Field Camera 3
Instrument Handbook \footnote{https://hst-docs.stsci.edu/wfc3ihb}. The net counts per beam were then converted to $\lambda$F$_{\lambda}$ using scaling parameters given by the inverse sensitivity keywords `PHOTFLAM' and the pivot wavelength `PHOTPLAM' with units ergs cm$^{-2}$ \AA$^{-1}$ counts$^{-1}$ and {\AA}~ respectively. 
Figure~\ref{fig:IR} and figure~\ref{fig:optical} display HST IR and optical images of our sources overlaid with the radio VLA image contours respectively. We do not present the IR image of PKS\,1036+054 which had poor image quality due to failure in guide star acquisition. 


\subsection{Radio data}
Three of the hybrid sources, namely, PKS\,0215+015, TXS\,0730+504, and PKS\,1036+054 were part of the Jansky VLA (JVLA) survey described in \cite{Stanley2017}. 
The details of the observations and image parameters are summarized in Table~\ref{tab:obs}.
Standard calibration and flagging procedures were performed using {\tt CASA} tasks. Automatic flagging tasks including \texttt{tfcrop} and \texttt{rflag} along with additional manual flagging were employed in removing RFI. After the basic calibration, these sources were imaged using the {\tt CASA} task {\tt tclean} while setting the parameters deconvolver=`mtmfs'  with two Taylor-series terms (nterms = 2). Five rounds of self-calibration were carried out, calibrating only the phase for the first two times followed by calibrating both amplitude and phase for three times. The self-calibration introduced phase errors, with negative artefacts in the A-array image of PKS\,1036+054 at 1.4~GHz. A narrower bandwidth was then used for the imaging and self-calibration rounds instead of the entire wide-band. Finally, we applied the time based gain solutions obtained to the entire bandwidth. The final image made with the entire bandwidth shows an improved image quality. Hence the phase errors might have been the result of the spectral variation of features like the lobe across the wide-band. 
The images observed using VLA B-array configuration at 5.5~GHz have matched resolution with those at 1.5~GHz observed in the A-array configuration. We make use of these images while estimating the extended flux densities from the jet regions which were used for SED modeling.

In addition to these data sets included in \cite{Stanley2017}, we also used available archival observations for PKS\,0215+015 using the JVLA (10~GHz) and ALMA (700~GHz). The target source was split out from the calibrated data sets that were downloaded from their respective archives and imaged using the MT-MFS algorithm with nterms=2 in {\tt CASA}. The JVLA image at 10~GHz resulted in detection of the jet feature, whereas the ALMA image did not show any jet-related feature. We used natural weighting while imaging the B-array data at 10~GHz to achieve a resolution similar to the 5~GHz B-array images. Three times the rms noise (3$\sigma$) from the ALMA image was used as an upper limit; this point was not included for carrying out the model fitting. Figures~\ref{fig:0215+015_vla}, \ref{fig5},  \ref{fig:0730+504_vla} and \ref{fig:1036+054_vla} show the radio images of the sources PKS\,0215+015, TXS\,0730+504 and PKS\,1036+054 respectively. The ALMA image of PKS\,0215+015 is not shown because the image does not show any details other than the central point source.

\section{Results}
\label{sec:results}

\begin{figure*}
 \centering
 \includegraphics[height=9cm,trim = 20 100 0 100]{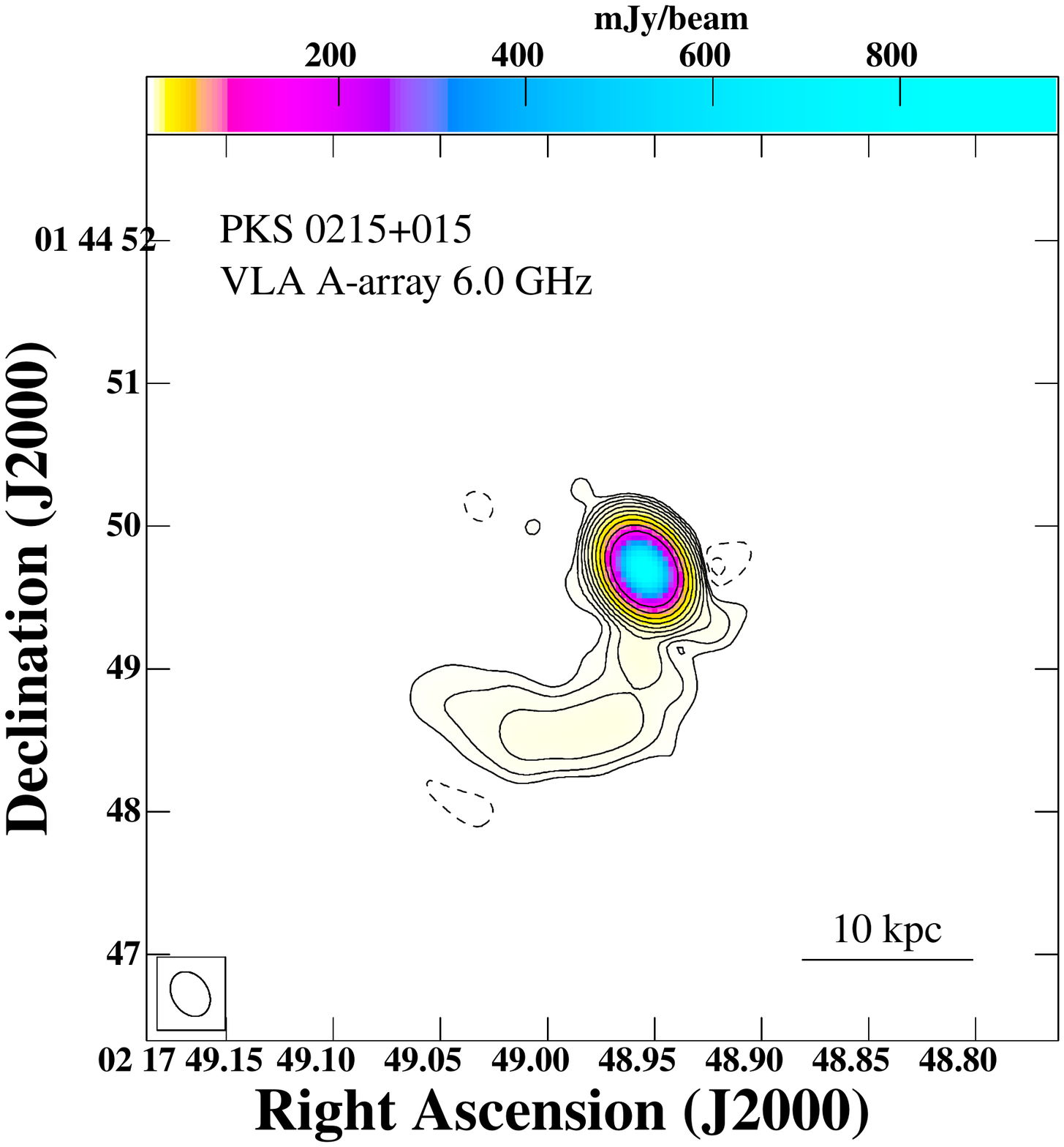}
 \includegraphics[height=9cm,trim = 20 110 0 112]{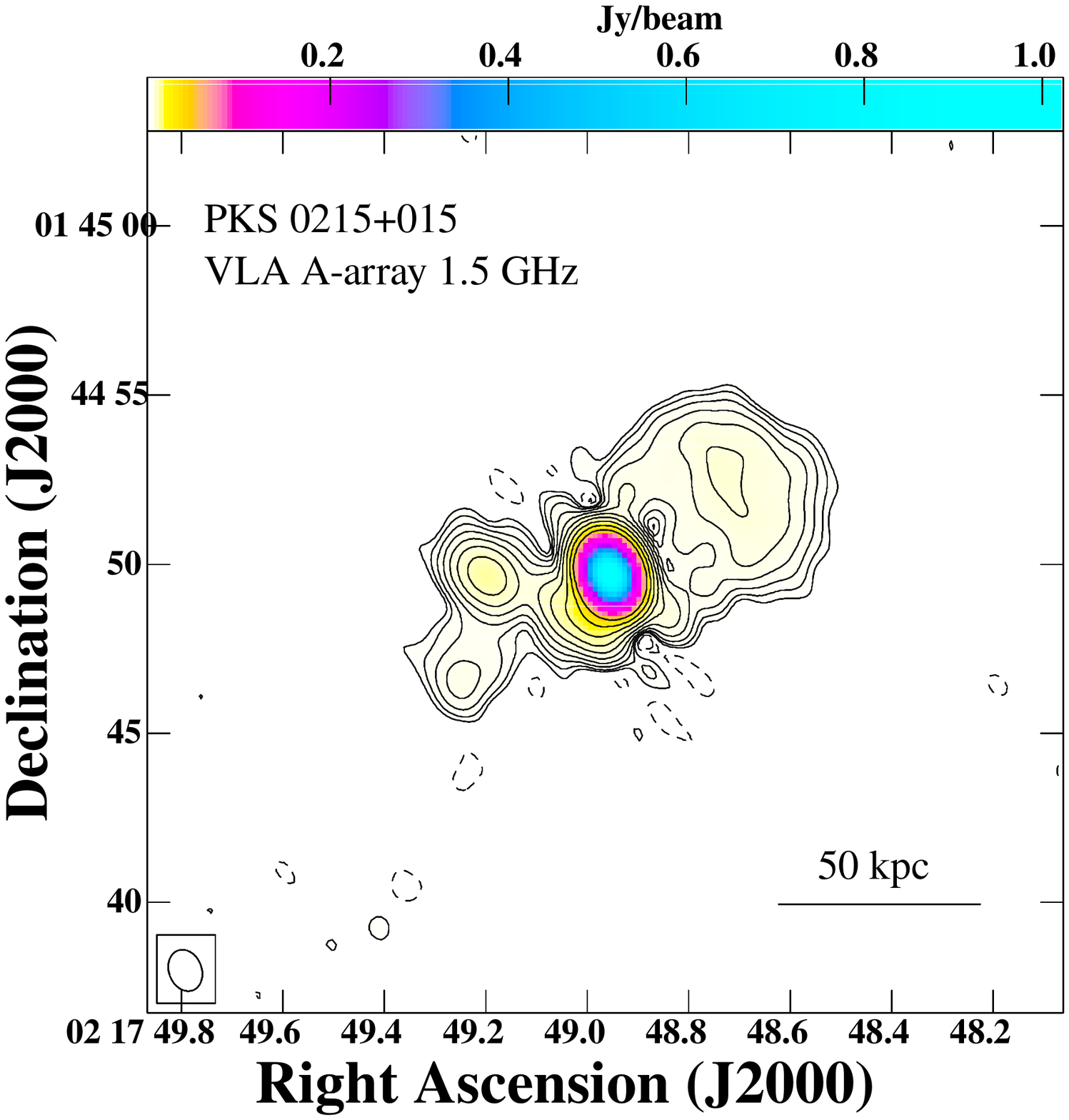}
 \includegraphics[height=9cm,trim = 20 100 0 100]{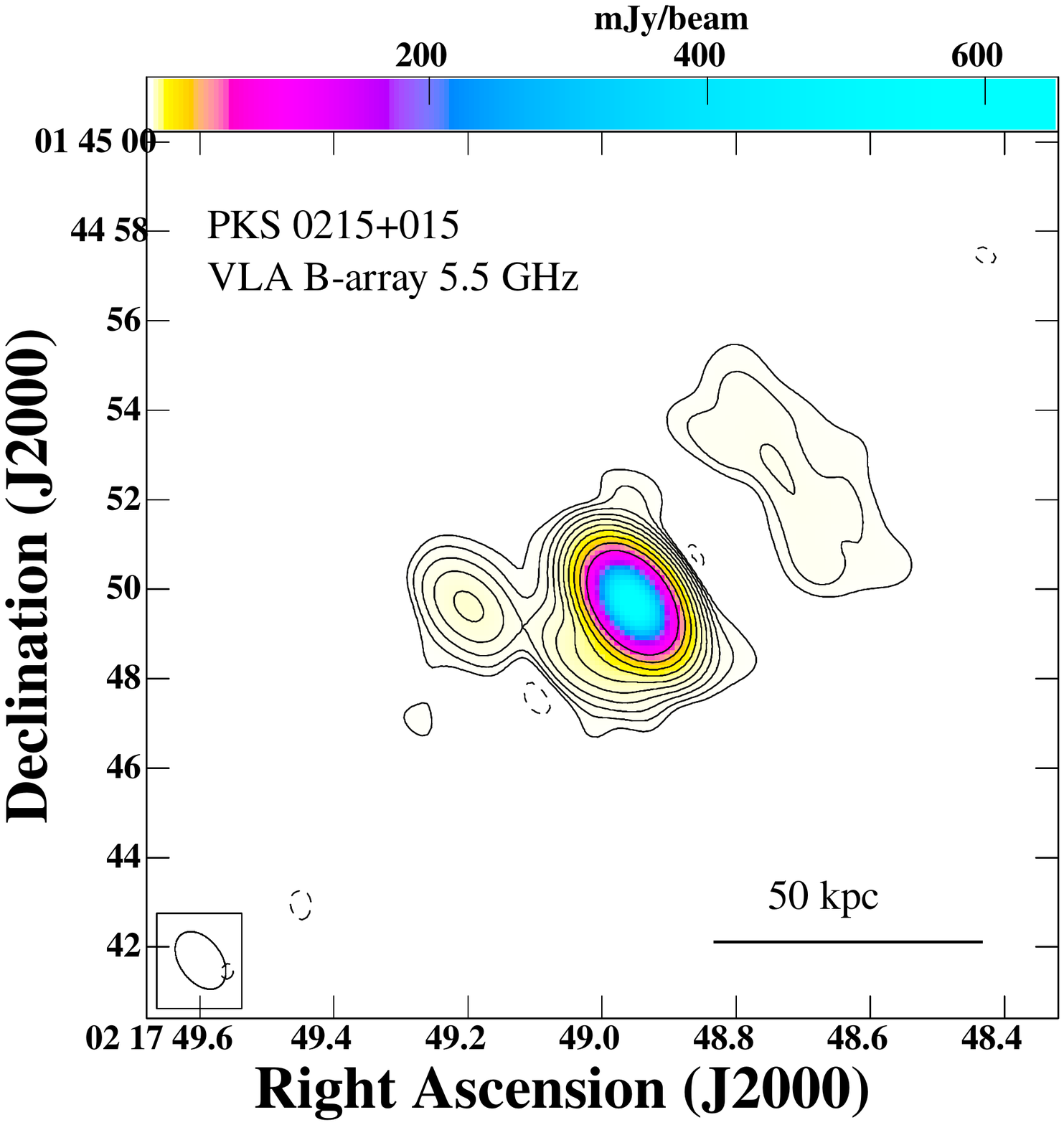}
 \includegraphics[height=9cm,trim = 20 100 0 100]{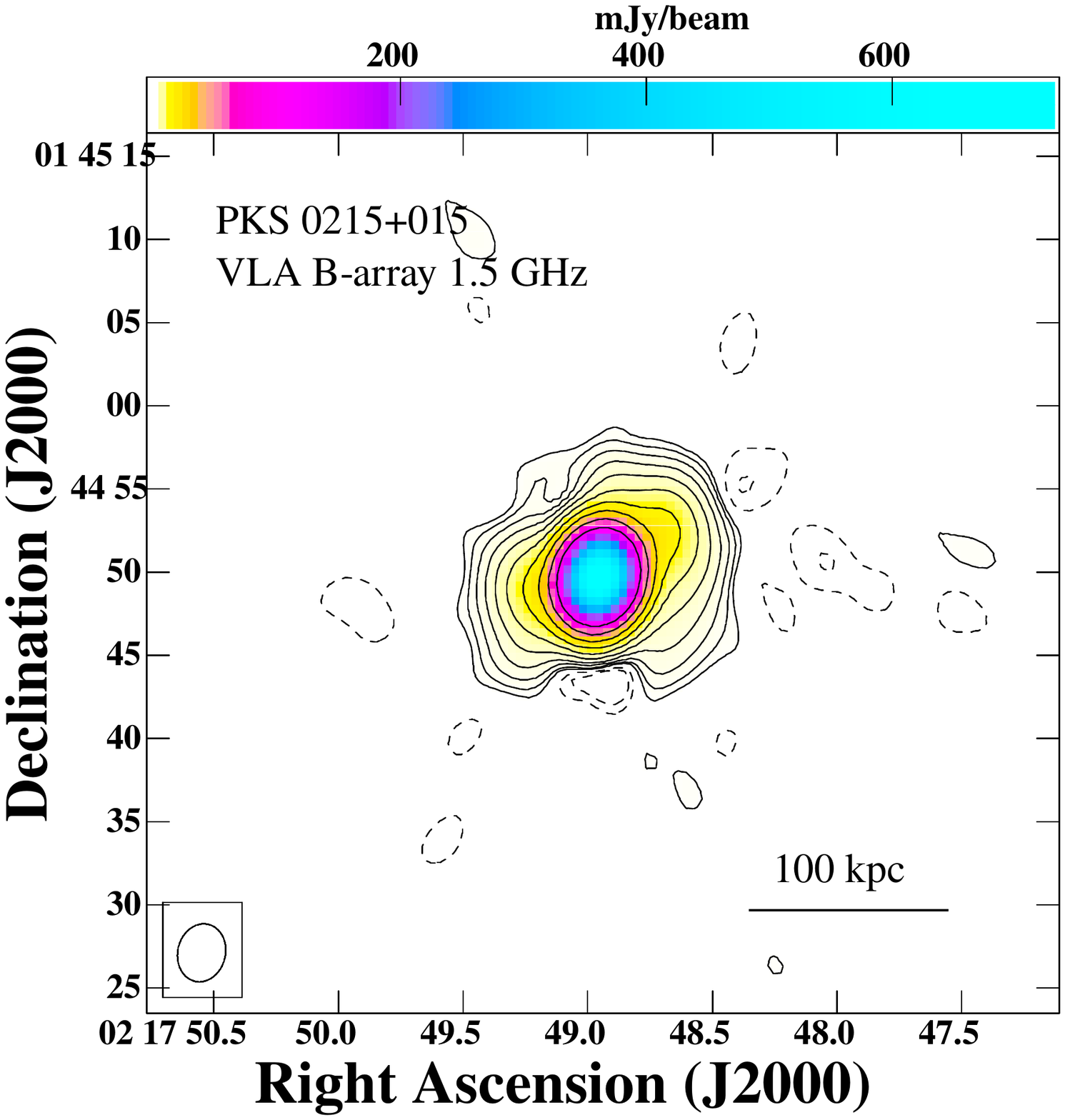}
 \caption{Radio total intensity images of PKS\,0215+015. The contour levels are chosen as 3$\sigma\times$(-2, -1, 1, 2, 4, 8, 16, 32, 64, 128, 256, 512), where $\sigma$ is the rms and is tabulated in Table~\ref{tab:obs} along with peak flux density and beam parameters. Top left: VLA A-array image at 6.0~GHz. Top right: VLA A-array image at 1.5~GHz. Bottom: VLA B-array image at 5.5~GHz. Bottom left: VLA B-array image at 1.5~GHz.}
 \label{fig:0215+015_vla}
\end{figure*}

\begin{figure*}
 \centering
 \includegraphics[height=10cm,trim = 20 100 0 100]{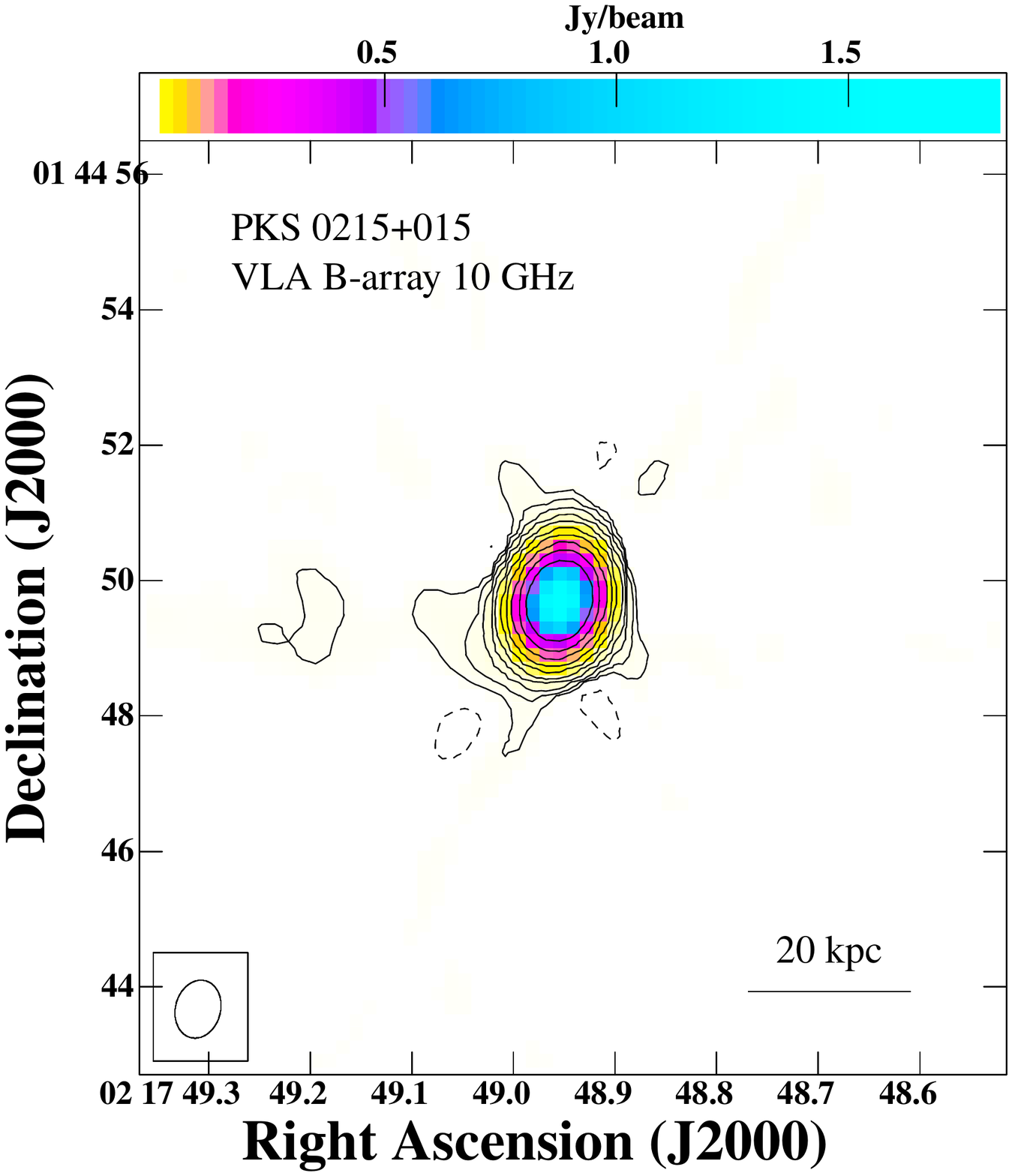}
 \caption{Continued --Radio total intensity images of PKS\,0215+015 at 10~GHz using B-array. The contour levels are chosen as 3$\sigma\times$(-2, -1, 1, 2, 4, 8, 16, 32, 64, 128, 256, 512), where $\sigma$ is the rms and is tabulated in Table~\ref{tab:obs} along with peak flux density and beam parameters.}
 \label{fig5}
\end{figure*}

\begin{figure*}
 \centering
 \includegraphics[height=18cm]{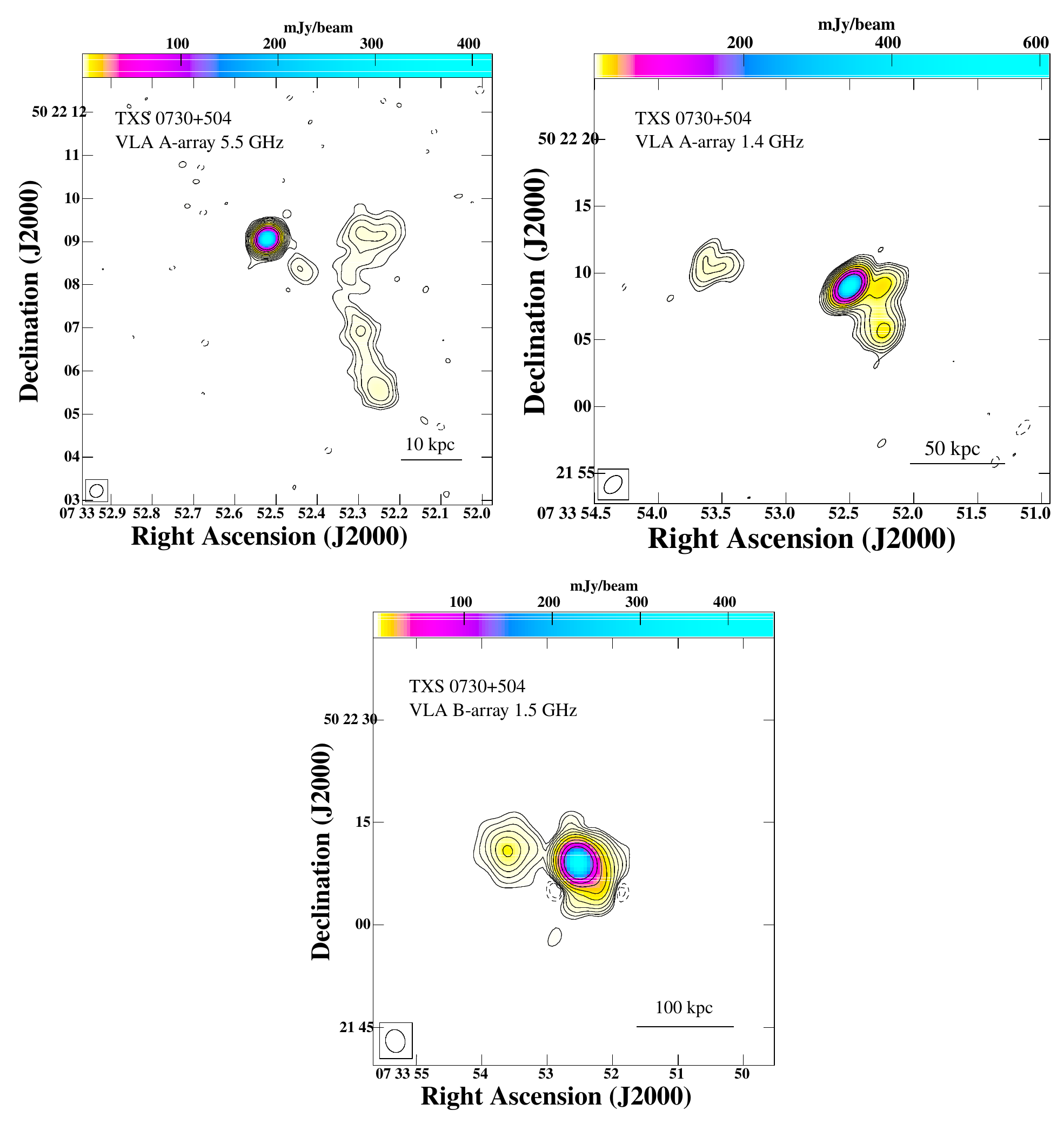}
 \caption{Radio total intensity images of TXS\,0730+504. The contour levels are chosen as 3$\sigma\times$(-2, -1, 1, 2, 4, 8, 16, 32, 64, 128, 256, 512), where $\sigma$ is the rms and is tabulated in Table~\ref{tab:obs} along with peak flux density and beam parameters. Top left: VLA A-array image at 5.5~GHz. Top right: VLA A-array image at 1.4~GHz. Bottom: VLA B-array image at 1.5~GHz.}
 \label{fig:0730+504_vla}
\end{figure*}
\begin{figure*}
 \centering
 \includegraphics[height=10cm,trim = 20 100 0 100]{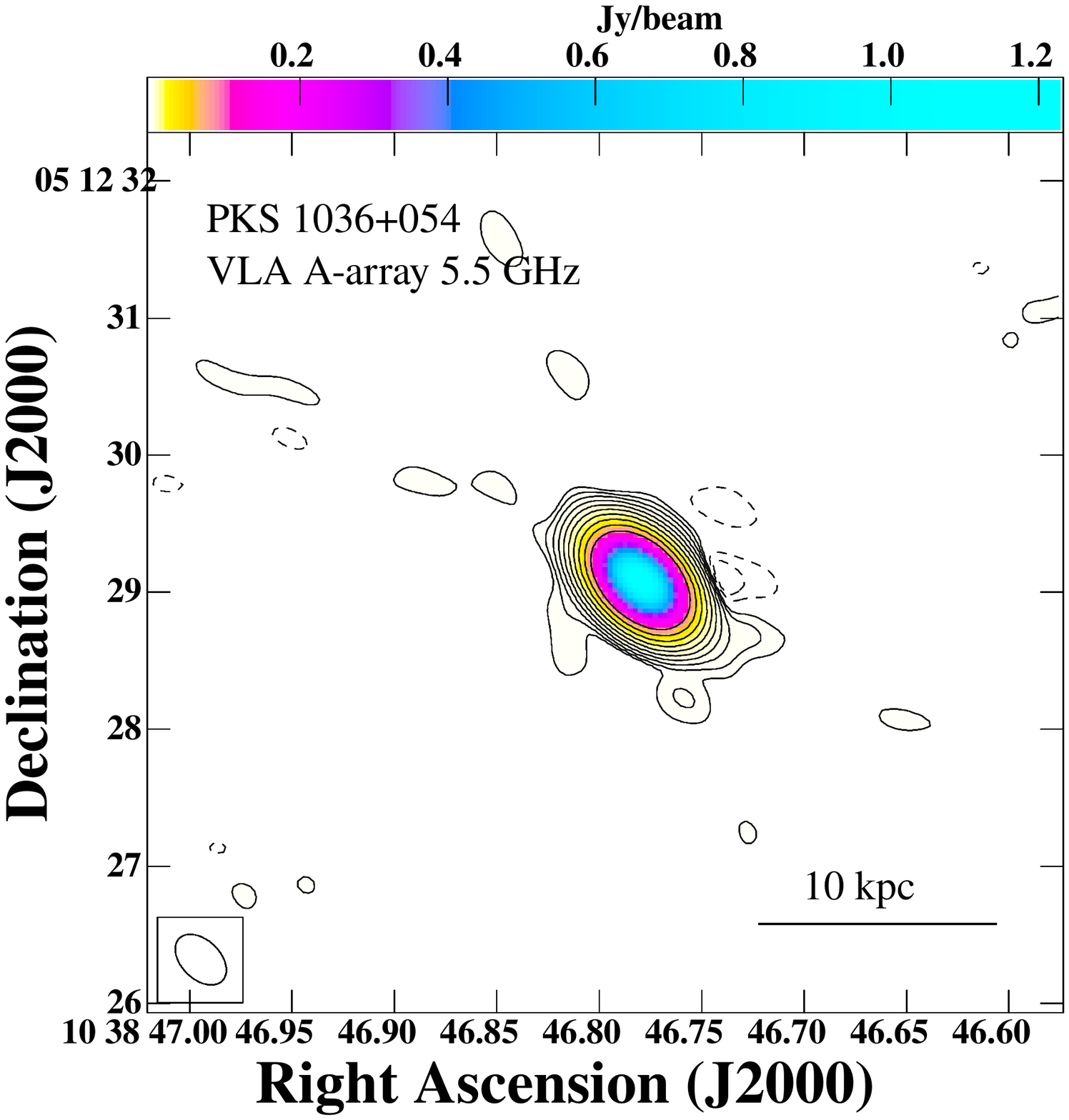}
 \includegraphics[height=9.8cm,trim = 20 100 100 100]{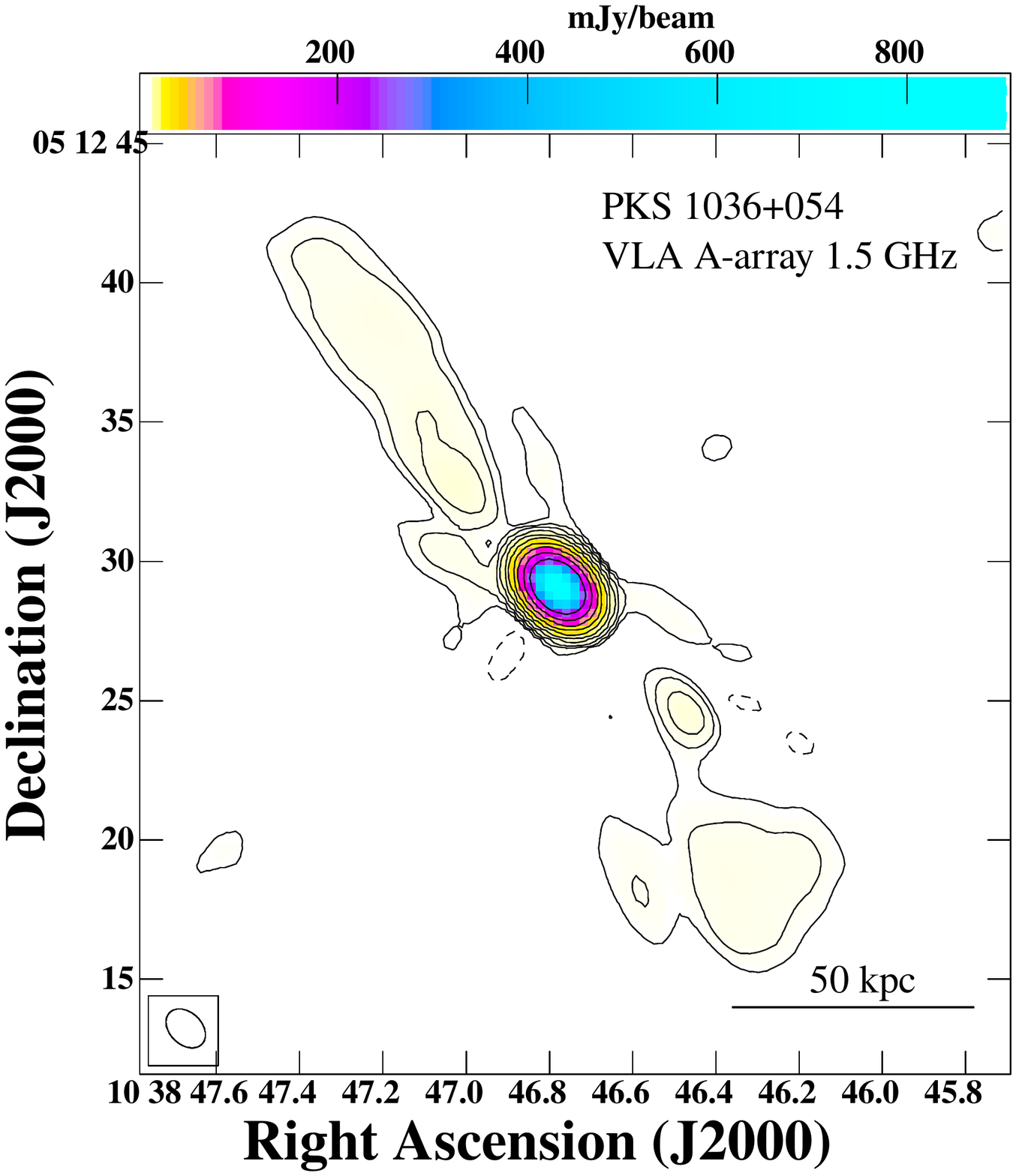}
 \includegraphics[height=10cm,trim = 20 100 0 100]{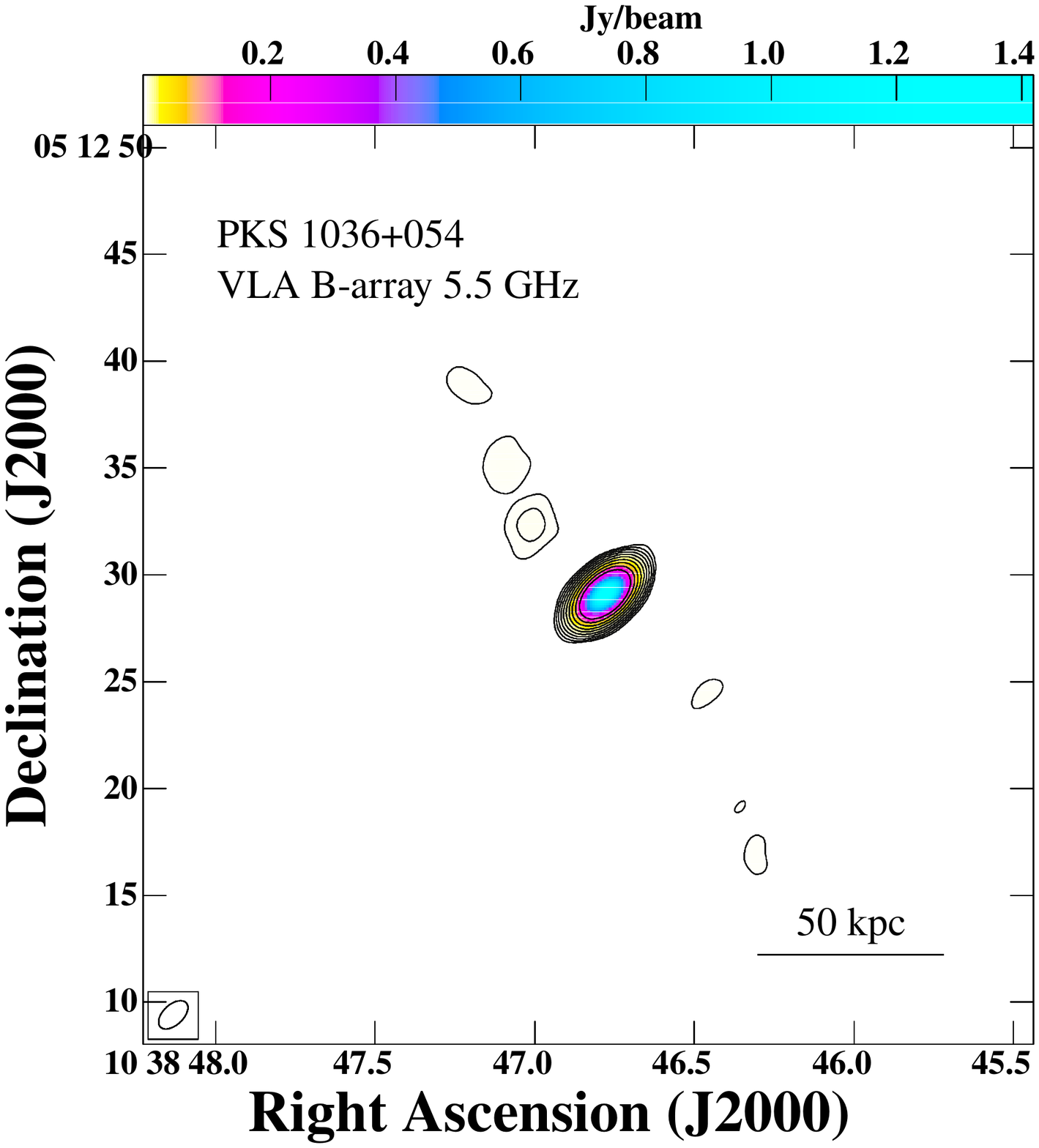}
 \includegraphics[height=10cm,trim = 20 100 0 100]{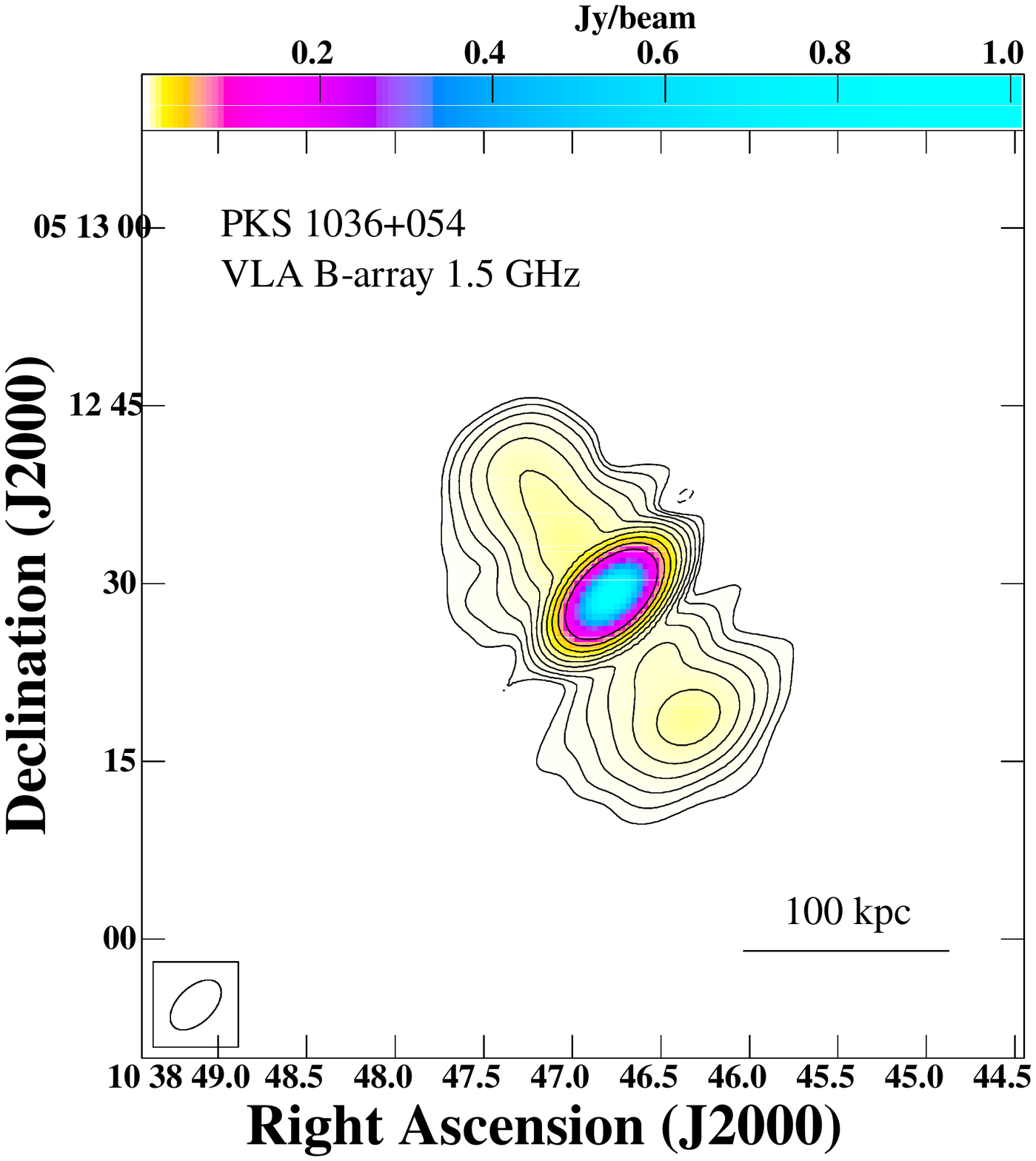}

 \caption{Radio total intensity images of PKS\,1036+054. The contour levels are chosen as 3$\sigma\times$(-2, -1, 1, 2, 4, 8, 16, 32, 64, 128, 256, 512), where $\sigma$ is the rms and is tabulated in Table~\ref{tab:obs} along with peak flux density and beam parameters. Top left: VLA A-array image at 5.5~GHz. Top right: VLA A-array image at 1.5~GHz. Bottom left: VLA B-array image at 5.5~GHz. Bottom left: VLA B-array image at 1.5~GHz.}
 \label{fig:1036+054_vla}
\end{figure*}

\subsection{Notes on individual sources}
\subsubsection{PKS\,0215+015}

PKS\,0215+015 shows large optical polarization \citep[17\%;][]{Wills1992} and \textit{Fermi}-LAT gamma-ray detection \citep{Abdollahi2020}. Radio observations reveal an inverted spectrum peaking above 10~GHz and extreme variability over time \citep{Torniainen2005}. VLBA observations display a one-sided jet in the west-east direction. While the lower resolution VLA images (figure~\ref{fig:0215+015_vla}) show lobe directions consistent with that of the VLBA jet, the VLA A-array image at C-band (see the top-left panel of figure~\ref{fig:0215+015_vla}) shows a bent jet that is pointed in the north-south direction near the core and is bent towards the east farther away. The apparent hotspot-like feature is resolved out in the highest resolution image, although the X-ray emission coincides with this brightness feature. 

\subsubsection{TXS 0730+504}
TXS\,0730+504 is a flat spectrum radio-quasar with a \textit{Fermi}-LAT $\gamma$-ray association \citep{Abdollahi2020}.
\cite{Kharb2010} presented the VLA 1.4~GHz image of TXS\,0730+504 which showed a hotspot towards the east and an FR\,I-like jet towards the west which then bends to the south abruptly. However, there seems to be a ``hotspot'' towards the southwest in the higher resolution image (see figure~\ref{fig:0730+504_vla}). This is also the source with the most pronounced X-ray emission that covers the entire jet and also coincide with the location of the hotspot. The VLBA pc-scale jet is also aligned in the same direction as the X-ray jet \citep{Pushkarev2017}. The inner jet in TXS\,0730+504 is pointing towards the hotspot and the rest of the emission appears diverted to the north, resembling wings in X-shaped radio galaxies \citep{Lal2019}. 

\subsubsection{PKS 1036+054}
PKS\,1036+054 is classified as a flat-spectrum radio quasar at a redshift of 0.473 \citep{Healey2007}. We do not detect significant amounts of X-ray emission from the extended jet regions on kpc scales. Unlike PKS\,0215+015 and TXS\,0730+504, PKS\,1036+054 does not show stark differences between jet directions in images with different resolutions. 

\subsubsection{PKS 1730-130}
PKS\,1730$-$130 is a highly variable FSRQ at a redshift of 0.9. We did not detect X-ray or optical emission from the extended jet regions of this source. The pc-scale \citep{Pushkarev2017} and the kiloparsec scale jets are misaligned from each other by an angle of  $\sim$90$\degr$.

\subsection{Spectral Energy Distribution Modeling}
\label{Sec:sedmodel}

\subsubsection{Fitting Method}
We carried out broadband SED modeling for the jet regions in the sources, PKS\,0215+015, and TXS\,0730+504, which showed X-ray jet detections using radio, optical and X-ray data points.
{ We used the optical data points despite being upper limits since these can provide crucial constraints on the SED models. We used 1$\sigma$ as the optical data point, where $\sigma$ is the noise level estimated from the respective optical images and used asymmetric error bars (+2$\sigma$, -1$\sigma$) to account for the uncertainties. Upon cross-matching with the {\textit{Fermi}} LAT fourth source catalog \citep[4FGL;][]{Abdollahi2020} we found that both PKS\,0215+015 and TXS\,0730+504 show gamma ray detections within the 95\% confidence radius. Since the gamma-ray emission is unresolved and may contain emission from the core along with the jet, we refrain from including these data points while carrying out the model fitting.
}

We used the {\tt agnpy}\footnote{\href{https://github.com/cosimoNigro/agnpy/}{https://github.com/cosimoNigro/agnpy/}} module \citep{Nigro2020} for the SED modeling. {\tt agnpy} is a python package that enables the modelling of the radiative processes in the jets of AGN. It uses Sherpa package \citep{Freeman2001} to carry out the fitting of the SED. We used the Chi-square statistic with the Gehrels variance\footnote{\href{https://sherpa.readthedocs.io/en/4.13.1/statistics/api/sherpa.stats.Chi2Gehrels.html}{https://sherpa.readthedocs.io/en/4.13.1/statistics/api/
sherpa.stats.Chi2Gehrels.html}} function and the Monte Carlo optimization\footnote{\href{https://sherpa.readthedocs.io/en/4.13.1/optimisers/api/sherpa.optmethods.optfcts.montecarlo.html}{https://sherpa.readthedocs.io/en/4.13.1/optimisers/api/
sherpa.optmethods.optfcts.montecarlo.html}} method which provided visually better fits. More details on the Monte-Carlo optimization method that we used can be found in \cite{Storn1995}
The model was chosen to be a sum of the synchrotron, synchrotron self-Compton (SSC), and IC-CMB emission. The underlying electron population was assumed to follow the same power law energy distribution for all three emission mechanisms. We assumed a single zone spherical blob model for the extended jet although the real jets may have more complex structures. We assumed this model to avoid increasing the complexity of the model and reduce the number of unknown parameters. The variable parameters of the model are the electron number density (n$_e$), power law index of the electron energy distribution (p), magnetic field (B), minimum and maximum electron Lorentz factors ($\gamma_{min}$ \& $\gamma_{max}$), the size of the emission region ($r$), the cosine of the viewing angle ($\mu_s$=cos($\theta$)) and the bulk Lorentz factor ($\Gamma$). We keep the minimum electron Lorentz factors frozen at $\gamma_{min}$=5.0 because varying it does not have a prominent effect on the SED. Similarly, $\gamma_{max}=1\times10^5$ was kept frozen while fitting the IC-CMB models since \cite{Stanley2015} have noted that this parameter has little effect on the SED modeling.

\begin{table*}[tbph]
\begin{center}
\caption{Best fitting values for the spectral energy density model parameters}
\begin{tabular}{cccccccccc} \hline \hline
%
Source & $n_e$ & p & $\gamma_{max}$ & B & $\theta$ & $\Gamma$& $\delta_D$ & $\beta$ & r \\
  & ($\times$10$^{-5}$cm$^{-3}$) & & ($\times10^5$) & ($\mu$G) & ($\degr$) & & & & ($\times$10$^{20}$cm) \\
(1) & (2) & (3) & (4) & (5) & (6)& (7) & (8) & (9) & (10)  \\
\hline

PKS\,0215+015 (SSC)$^a$ & 1.5 & 1 & 59.6 & 0.37 & 25.8 & 5.6 & 1.56 & 0.984 & 26.5 \\
PKS\,0215+015 (IC-CMB)$^b$  & 1.0 & 2.15 &  1$^*$ & 55.0$^*$&0.6
$^*$& 37.4$^*$ & 65.0$^*$ & 0.999$^*$&89.8\\
TXS\,0730+504 & 3.78& 1.90 & 1$^*$& 3.16& 2.00& 13.5 & 22.2 & 0.997& 7.72 \\
\hline
\label{tab:parameters}
\end{tabular}
\\{$^*$- parameters that were kept frozen, $^{a,b}$- Fit parameters corresponding to the top and bottom panels of figure~\ref{fig:sed:0215} respectively. 
n$_e$: electron number density, p: power law index of the electron energy distribution following N(E)=E$^{-\mathrm{p}}$, $\gamma_{max}$: maximum electron Lorentz factors, B: magnetic field, $\theta$: the viewing angle ($\theta$=cos$^{-1}$($\mu_s$), $\Gamma$: the bulk Lorentz factor, $\delta_{D}$: the Doppler factor, $\beta$: ratio of jet velocity to velocity of light (v/c), $r$: the size of the emission region}
\end{center}
\end{table*}
One of the main challenges that various models face despite being able to reproduce the SED is that the parameters predicted by the best fit models are not consistent with the jet region properties deduced from other observational constraints. We limit the parameters to a physically viable space which were obtained using other methods and we discuss these in the following subsections.

\subsubsection{Parameter constraints}
\setcounter{secnumdepth}{4}
\paragraph{Size of emission region:}

Although varying the emission region size ($r$) does not substantially alter the overall shape of the SED, the size affects the absolute value of the flux. Consequently, it affects the other parameters that determine the absolute flux levels, including the magnetic fields, and electron number density while finding the best fit model. This degeneracy between several of the parameters makes it important to constrain the emission region size, if possible, using alternative methods. The X-ray emission from TXS\,0730+504 appears resolved and spans the length of the jet as can be inferred from figure~\ref{fig:x-ray}. In PKS\,0215+015, it is hard to discern whether the emission from the hotspot-like feature is resolved due to the low SNR of the X-ray emission from the jets. The hotspot-like feature is absent in the VLA A-array image at 5.5~GHz which has the highest resolution in our study, suggesting that the emission region size is higher than the resolution of this image. We fit a Gaussian model to the `bright knot' using the `{\tt JMFIT}' task in {\tt AIPS} from the B-array image at 5.5~GHz, which is at a coarser resolution. The source appears resolved from this exercise of Gaussian fitting. In addition, we assumed that the peak flux density estimated from the 5.5~GHz B-array image was distributed uniformly in the A-array image within a region of size equal to one resolution element of the B-array image. The noise in the A-array would lead to a non-detection in A-array image if the flux density is uniformly distributed. Hence we conclude that the non-detection in the A-array image is due to the diffuse nature of the emission.
The emission regions in both PKS\,0215+015 and TXS\,0730+504 are resolved in our highest resolution radio images.
Hence we allow `$r$' to vary between the size corresponding to the highest resolution 
and the radius of the region chosen to estimate the flux using DS9. The ranges are 8.7$\times$10$^{21}$cm - 4.9$\times$10$^{22}$~cm and 4.3$\times$10$^{21}$cm - 5.1$\times$10$^{22}$~cm for PKS\,0215+015 and TXS\,0730+504 respectively.
\paragraph{Jet Kinematics}
\label{sec:jetkine}
VLBA data provide additional constraints on the jet kinematics at pc scales. While the jets could have decelerated significantly from pc to kpc scales, the pc-scale apparent jet velocity is a firm upper limit. For PKS\,0215+015, in addition to $\beta_{\rm{app}}$, information about the Doppler factor ($\delta_{\mathrm{D}}$) is also available \citep[$\delta_D \sim 65.0$][]{Homan2021ApJ}. Hence, we can determine the viewing angle, $\theta$=cos$^{-1}(
\mu_s)$ and $\beta$ by using the following equations.
$$
\theta = \arctan{\left( 2\beta_\mathrm{app} \over
\beta_\mathrm{app}^2+\delta^2-1\right)}
$$
$$
\Gamma =  {(\beta_\mathrm{app}^2 + \delta^2 +1) \over 2\delta} 
$$
where $\Gamma=\sqrt{\frac{1}{1-\beta^2}}$ is the bulk Lorentz factor.
%
We assumed that the jet could only have undergone deceleration from pc to kpc scales and an upper limit was imposed on $\beta_{\rm{app}}$ while carrying out the fitting.
\paragraph{Equipartition Magnetic fields}
 
 We estimated equipartition magnetic fields \citep{Burbidge1959} assuming a spherical geometry with an emission region size of 10$^{22}$cm which is of the same order as that of the DS9 region chosen to estimate the flux density for SED modeling. We carry out an order of magnitude estimation since there are several unknowns. For example, the emission region size in the jet frame will be scaled by a factor of 1/$\Gamma$, and the electron population may not be uniformly distributed in the entire region.  
 It was shown by \cite{Hardcastle2004} and \cite{Croston2005} that the magnetic fields estimated from the lobes/hotspots of FR\,II sources seemed to be off by less than an order of magnitude from the equipartition values.
Since it is not known whether equipartition assumption holds for jet knots as well and the estimation of the equipartition magnetic fields has high uncertainties, we allow the magnetic fields to vary within a two orders of magnitude range centered on the equipartition magnetic fields while carrying out the SED model fitting. 

\paragraph{The case of PKS 0215+015}
\label{sec:0215_case}
As already mentioned in  Section~\ref{sec:jetkine}, the jet on pc scales in PKS\,0215+015 has constraints on both the viewing angle and jet velocity. We assumed that the viewing angle remains the same from pc to kpc scales whereas the jet speed might have undergone deceleration. Hence, the parameter $\mu_s$ was frozen and an upper limit was imposed on $\beta_{\rm{app}}$ while carrying out the fitting.   We also note that the low SNR of the X-ray data points lead to large errors on the derived power law parameters (i.e., photon index $=0.5 \pm 0.8$ and normalization $=6.0 \times 10^{-7} \pm 5.7 \times 10^{-7}$; see Section~\ref{sec:chandra_dataanal} for more details) which were in turn used to estimate the X-ray points for the SED modeling. These stringent constraints along with the low SNR of the X-ray data points rendered it difficult to find a best fit model that matches the data well. Hence, we do not present this SED modeling carried out using the broadband data from radio through X-ray wavelengths for PKS\,0215+015 in this paper.

Instead, we fit the data (only including HST and radio data) using a simple synchrotron model under the assumption that synchrotron mechanism is the single major contributor to the emission in radio-optical wavebands. The results of this exercise is shown in the upper panel of  Figure~\ref{fig:sed:0215}. Several parameters were kept frozen including magnetic fields (see Table~\ref{tab:parameters} for details).
Once the parameters of the underlying electron energy distribution were determined, we calculated the IC-CMB and SSC emission produced by the same electron population for the given $\mu_s$ and $\beta_{\rm{app}}$. This model is consistent with the ALMA and the \textit{Fermi} upper limits (see Section~\ref{sec:gamma-ray-emission}). While the X-ray slope predicted by the model is slightly off from the measured value, the model is consistent with most of the X-ray data points within the error bars. 


In addition to the above fitting, we also tried relaxing the stringent constraints on all parameters to obtain a best fit model and is illustrated in the bottom panel of Figure~\ref{fig:sed:0215}.

We note that there are a few caveats that limit the scope of our results from SED model fitting.
The SED lacks flux detections at frequencies in between the radio and X-ray frequencies. Since HST data points are upper limits, the constraining of the low-frequency peak, the subsequent dip and the rise are poorly constrained. Another factor is the large number of parameters in the SED model, which makes the set of estimated best fit parameters one of many possible combinations. 
\subsubsection{Results}%
\label{sec:result:sub}

\begin{figure*}
 \centering
 \includegraphics[height=9cm,trim = 0 0 0 0 ]{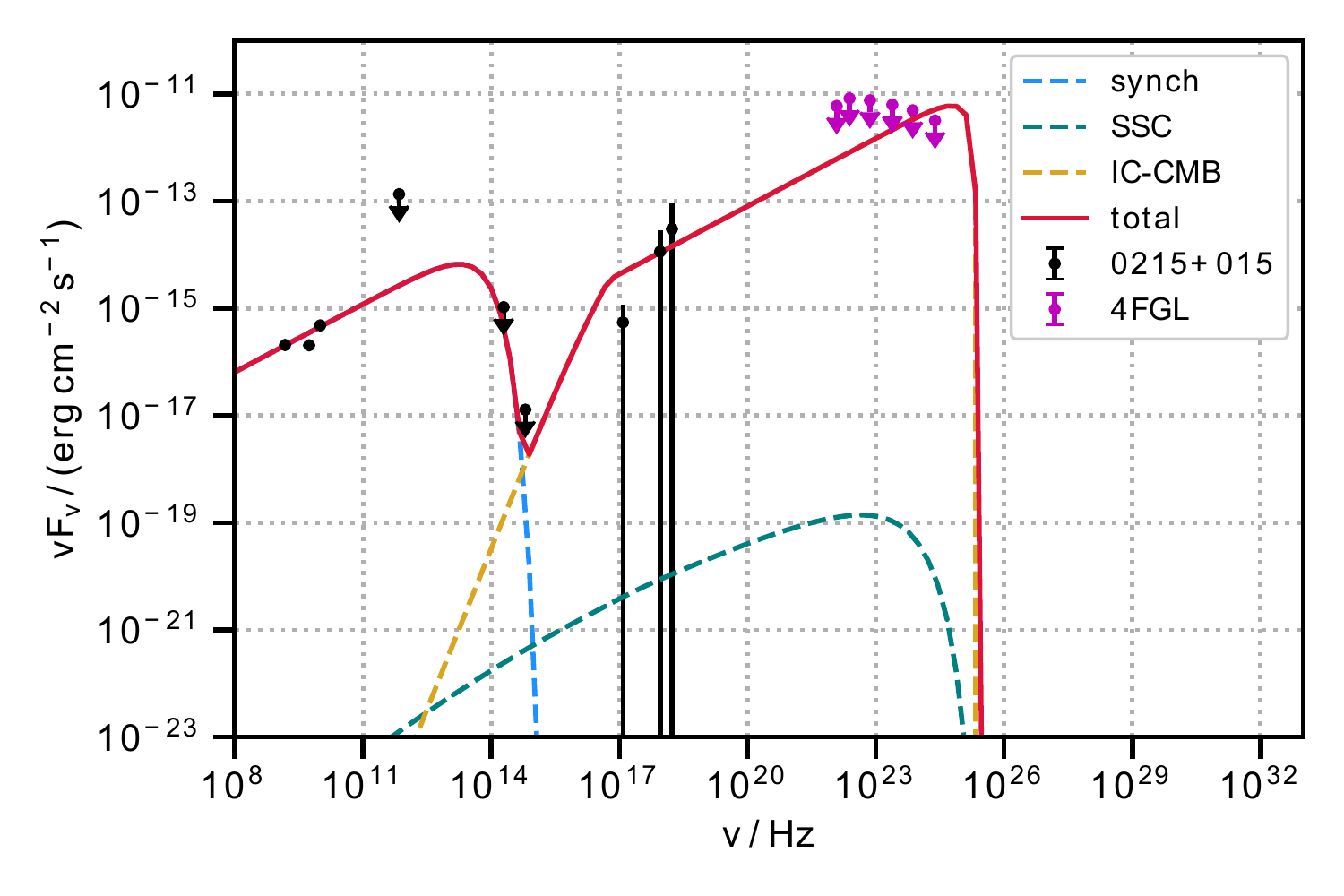}
 \includegraphics[height=9cm,trim = 0 0 0 0 ]{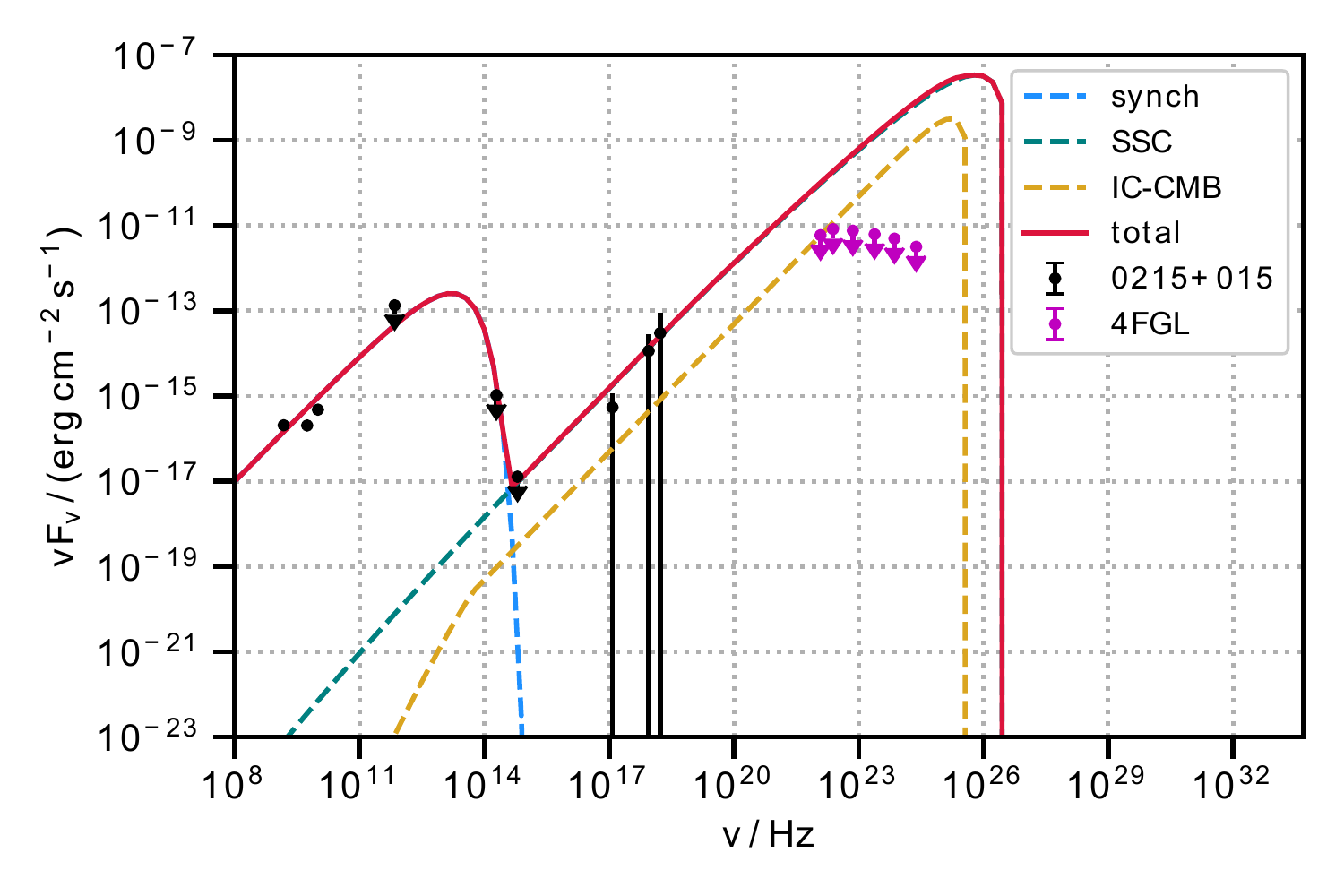}
 \caption{Broadband SED models of extended jet regions of PKS\,0215+015 with parameters estimated by fitting radio to optical data points with a single synchrotron model and using these derived parameters to calculate the IC-CMB emission at higher energies (top panel) and relaxed parameter space (bottom panel; see Section~\ref{sec:result:sub} for details). The black data points correspond to the emission from the extended jet region obtained from the VLA, ALMA, HST, and \textit{Chandra} images. The magenta upper limit points represent the gamma-ray flux from the 4FGL source catalog and were not used for fitting. The blue dashed line represents the synchrotron emission model, whereas the green and yellow dashed lines represent the SSC and IC-CMB models, respectively. The total emission from all the different components is depicted by the red solid line. }
 \label{fig:sed:0215}
\end{figure*}

\begin{figure*}
 \centering
 \includegraphics[height=9cm,trim = 0 0 0 0 ]{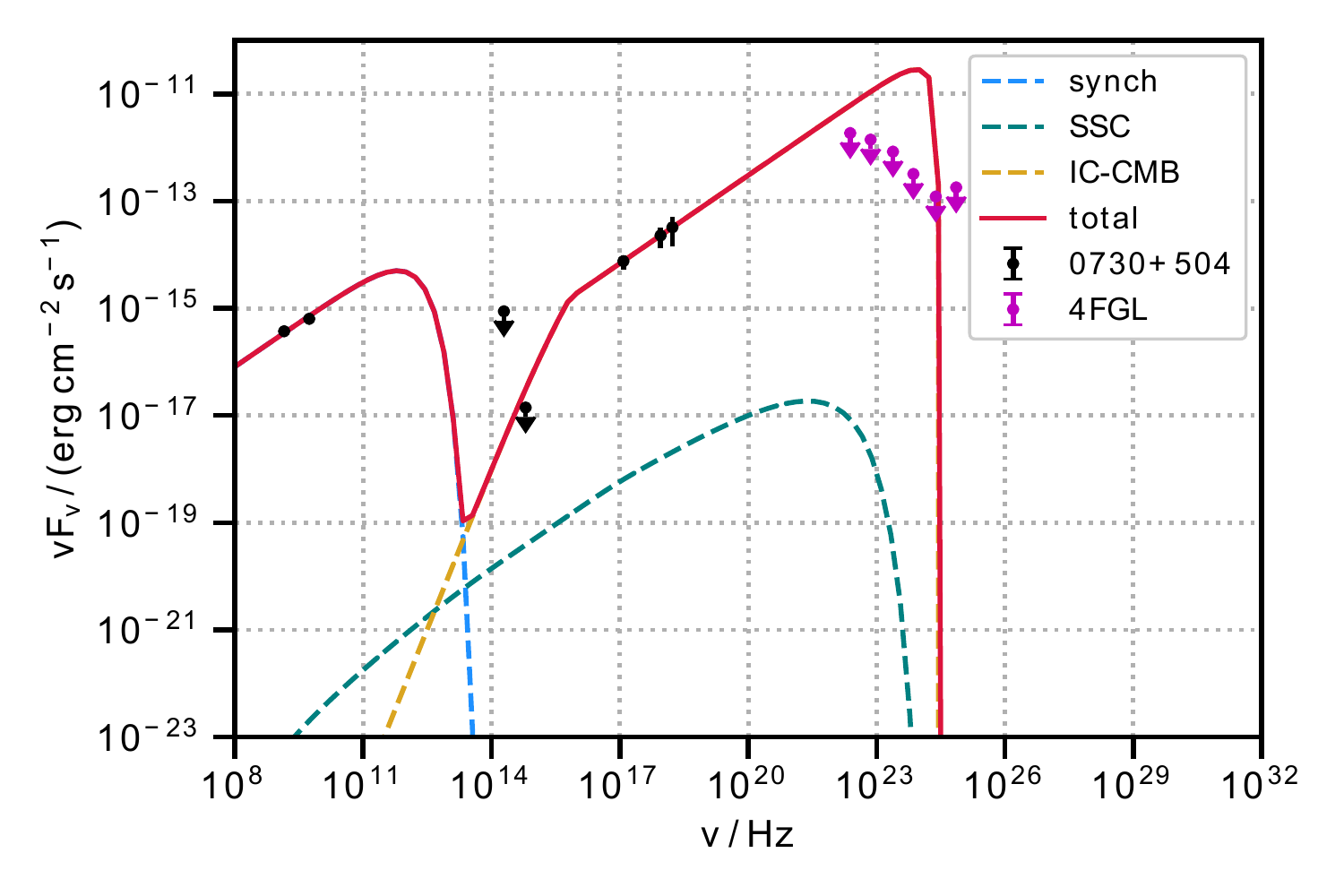}
 \caption{Broadband SED models of extended jet regions of TXS\,0730+504.The black data points correspond to the emission from the extended jet region obtained from the VLA, HST, and \textit{Chandra} images. The magenta upper limit points represent the gamma-ray flux from the 4FGL source catalog and were not used for fitting. The blue dashed line represents the synchrotron emission model, whereas the green and yellow dashed lines represent the SSC and IC-CMB models, respectively. The total emission from all the different components is depicted by the red solid line.}
 \label{fig:sed:0730}
\end{figure*}

The SED of the extended jet from PKS\,0215+015 and TXS\,0730+504 are shown in figure~\ref{fig:sed:0215} and \ref{fig:sed:0730}. The top panel in figure~\ref{fig:sed:0215} shows the case where we have fit the data points from radio to optical wavelengths with a synchrotron model and calculated the resulting IC-CMB and the SSC emission from the same population of the electrons, whereas the bottom panel represents the case where we have relaxed the stringent constraints on all parameters.
The non-detection of the jet in optical frequencies instantly rules out the possibility of a single synchrotron model (we refer to a single population of electrons whose electron energy distribution
decreases monotonically at higher energies) being able to explain the entire spectrum. 

The bottom panel of figure~\ref{fig:sed:0215} shows the case where SSC model fits the X-ray emission from the jet of PKS\,0215+015 better than the IC-CMB model. The value of $\Gamma$ complies with that expected from the jet-to-counter jet ratio of the X-ray emission.
However, the magnetic fields are less than the equipartition value by around two orders of magnitude (see Table~\ref{tab:parameters}) similar to previous studies \citep{Harris2006}. Moreover, the $\Gamma$-ray emission is also over predicted by the SSC model. Hence the SSC model is unlikely to be the major contributor to the excess X-ray emission seen in these jet regions and can be ruled out.

In figure~\ref{fig:sed:0730} and the upper panel of figure~\ref{fig:sed:0215}, the IC-CMB model dominates the total high-frequency emission in both sources, when stricter constraints are imposed on the parameters as discussed in the previous section. 

\section{Discussion}
\label{sec:discussion}
\subsection{Feasibility of the IC-CMB model}
Although the IC-CMB model seems to give better fits to the X-ray data in both the sources several issues with IC-CMB model were pointed out in the literature including highly relativistic jet speeds on kpc-scales or very small beaming angles requiring unrealistically large de-projected jet lengths \citep{Dermer2004,Sambruna2008}, over-prediction of $\gamma$-ray flux levels \citep{Meyer2014} and, super-Eddington jet kinetic power \citep{Dermer2004,Uchiyama2006}. In the following section, we investigate these factors and whether the IC-CMB model can overcome the above-mentioned challenges for the two sources in our sample.

\subsubsection{Jet kinetic energy}
We calculate the jet kinetic energy using the following relation from \cite{Celotti1997}.
\begin{equation}
 L_j=\pi r^2 n_e m_e c^2 \Gamma^2 \beta c \,\,\,\mathrm{ergs\,\,s^{-1}}
\label{eq1}
\end{equation}
{Several assumptions underlie these calculations \citep[see][for details]{Celotti1997,Celotti1993}. 
For example, the matter content and the electron energy distribution are uncertain. Equation~\ref{eq1} assumes either an electron-proton plasma with $\gamma_{min} \approx$ 100 or an electron-positron pair plasma with $\gamma_{min} \approx$ 1.}
By employing the parameters derived from our data (see Table~\ref{tab:parameters}), we estimated L$_\mathrm{j}$ = 7.6$\times$10$^{43}$ ergs s$^{-1}$ and 3.2$\times$10$^{43}$ ergs s$^{-1}$ for PKS\,0215+015 and TXS\,0730+504 respectively. The jet kinetic energy derived for the sources in our sample are smaller than typical values for L$_\mathrm{j}$ which fall in the range 10$^{44}-$10$^{49}$ ergs\,s$^{-1}$ \citep{Celotti1997}. The X-ray luminosities, L$_\mathrm{x}$ for PKS\,0215+015 and TXS\,0730+504 are $\approx$ 10$^{47}$ and 10$^{45}$ ergs\,s$^{-1}$ respectively. We have assumed that the major contributor to the X-ray luminosity is accretion of matter on to the supermassive black hole of the galaxy. We hence use the X-ray luminosity as a proxy for the radiative output from accretion. Hence, these jets do not require super-Eddington kinetic powers and can comfortably be powered by standard central engines.
However, it is possible that our assumption of the X-ray emission being accretion dominated is incorrect and the jet is the major contributor to the X-ray emission. In this case, we will need to disentangle the contribution from both the components before concluding that the jet kinetic powers are not super-Eddington.

\subsubsection{Gamma-ray emission}
\label{sec:gamma-ray-emission}
With the launch of {\textit{Fermi}}, it became possible to constrain the upper end of the SED. While the coarse resolution of {\textit{Fermi}} LAT does not allow spectral modeling of various components such as the core and the jet separately, the data provide an upper limit to emission from the jet. 

In PKS\,0215+015, the gamma-ray flux points lie above the SED obtained (figure~\ref{fig:sed:0215} (top panel)). It has to be noted that the gamma-ray fluxes are not the low-state fluxes, and the SED of PKS\,0215+015 need not comply with stricter upper limits.
On the other hand, the SED of TXS\,0730+504 over predicts the gamma-ray emission that is observed (figure~\ref{fig:sed:0730}). The data coverage is not complete in the radio wavelengths and since the peak frequency of the synchrotron emission is not well constrained, the current fit is not sufficient to rule out the IC-CMB model. The gamma ray peak location is given by $\nu_{peak} \approx \nu_{cmb}\delta_D^2 \gamma_{max}^2$ \citep{Tavecchio2000}. $\nu_{cmb}$ is well known but $\gamma_{max}$ is not that well constrained and a typical value derived for other sources was used.
Hence, more data are required to conclusively rule out the IC-CMB model for this source.
\subsubsection{Physical extents, beaming angle and Lorentz factor}
\label{phys_extent}
From the SED modeling, we derived the beaming angle ($\theta$) and the Lorentz factor ($\Gamma$). For the specific case of PKS\,0215+015, the beaming angle and the Lorentz factors were kept the same as that derived from the VLBI observations since any decrease in speed will result in lower IC-CMB flux, and hence unable to explain the X-ray emission seen in this system (see Section~\ref{sec:results} for more details). The jet speeds on the kpc scale corresponding to the $\Gamma$ and $\theta$ was frozen at 0.999c for PKS\,0215+015 and was derived to be 0.997c for TXS\,0730+504 from the best-fit model (see Table~\ref{tab:parameters}). 
Several studies exploring the speeds of jets on kpc scales have often reported much smaller velocities. These Lorentz factors are unrealistically high, based on other studies of kpc-scale jets in a general population of radio galaxies  \citep{Wardle1997,Laing2002,Arshakian2004}.

Secondly, we estimated the deprojected linear size, given the beaming angle of our sources. The length from core to the farthest end of PKS\,0215+015 and TXS\,0730+504 are 2.3 Mpc and 2.4 Mpc respectively, making these some of the largest known sources in the literature \citep{Dabhade2020}.
The rarity of such large giant radio galaxies and the unrealistic requirement for the jets at such large distances to be relativistic pose additional problems for the IC-CMB model.

On the other hand, the small viewing angle of the X-ray jet does not always require large sizes if the jet is undergoing bending. The VLBI pc-scale jet \citep{Lister2019} and the X-ray jet in both TXS\,0730+504 and PKS\,0215+015 have similar position angles. Nonetheless, there is evidence for significant jet bending in PKS\,0215+015 and minor bending in TXS\,0730+504 at intermediate resolutions (see figure~\ref{fig:0215+015_vla} and figure~\ref{fig:0730+504_vla}). The projected sizes are 55~kpc and 35~kpc for PKS\,0215+015 and TXS\,0730+504 respectively. So even in the most extreme case of jet bending, the jets still need to be relativistic at kpc-scales. 

Another argument in favor of the IC-CMB model is the similarity in the spectral slopes of the X-ray and radio emission in TXS\,0730+504 \citep{Georganopoulos2006,Meyer2014}. The large X-ray measurement errors of PKS\,0215+015 make it hard to determine whether this is true for this source as well.

Hence, while the IC-CMB models are faced with severe challenges, there are ways to reconcile the data to this model. Only with additional data coverage and information about the polarization at different wavebands can the IC-CMB model be completely ruled out.
In the following section, we explore alternate models, which could explain the X-ray emission without having to depend on extreme jet speeds or viewing angles.

\subsection{Alternate mechanisms of origin}
A multiple-component synchrotron emission model is a viable alternative 
\citep{Hardcastle2006,Uchiyama2006,Cara2013} that favors the synchrotron origin over the IC-CMB model due to the high degrees of polarization in the optical and UV wavelengths.
However, the similarity in spectral slopes of the X-ray emission and radio emission in  TXS\,0730+504 is probably not a coincidence and indicates that the electron populations must be correlated \citep{Schwartz2000}.
The X-ray emission seen in TXS\,0730+504 appears nearly continuous along the jet as is also the case with many others in the literature, whereas the lifetime of an X-ray emitting electron population is only a few years \citep{Harris2006}.
Hence if the X-ray emission indeed has a synchrotron origin, the electrons must be undergoing in-situ acceleration. 

Shear layer acceleration or stochastic acceleration due to a turbulent layer surrounding the inner-jet has been proposed as a possible explanation \citep{Stawarz2002}. \cite{Tavecchio2021} further explore this possibility and recover reasonable values for magnetic fields and beaming parameters. 
Future investigation with radio polarization observations that resolve the transverse jet can shed some light on this model of acceleration.

Another possible explanation is external Compton emission from a different photon field other than the CMB. Given that the X-ray jets are located at a minimum distance of several tens of kpcs from the core in both our hybrid sources, it is unlikely that the accretion disk or other emission from the galaxy could form a photon source.

\subsection{Dualism based on FR morphology}
Previous studies of hybrid morphology AGN have shown that the majority of the X-ray jets in high-power systems can be explained using IC-CMB models. \cite{Stanley2015} note that the X-ray jets in low radio-power hybrid systems do not require IC-CMB models and can be satisfactorily explained using single-zone synchrotron models which suggests that the trend seen based on the FR morphology is determined by the radio power rather than by the morphology itself.

We used the 74~MHz flux densities and a typical spectral index value of $-0.7$ to estimate the flux densities at 1.4~GHz. Lower frequency observations recover more diffuse emission and hence reduce the effect of beaming to some extent. 
Both the sources in our sample have ``high'' radio powers and require an IC-CMB model or other alternatives to explain the X-ray emission in the jets.

It has been suggested that the origin of hybrid morphology sources is due to the differences in the environment \citep{Gopal-Krishna2000}. While the difference between the X-ray count rates of the region around the FR\,I and FR\,II lobes in PKS\,0215+015, TXS\,0730+504 and PKS\,1036+054 are less than 1$\sigma$, that in PKS\,1730-130 is as high as $\sim$2.5$\sigma$. More interestingly, it is the FR\,II type lobe that resides in the denser environment. This result contradicts our current understanding according to which we expect the FR\,I jets to lose their collimation and stability early on due to the denser environment. We find that this asymmetry in X-ray counts is also present while comparing the inner sectors. Although we only included regions beyond the jet based on the radio images to estimate the asymmetries, some diffuse emission below the sensitivity of our observations is likely to be present in the outer regions. Hence, an alternate explanation is that the jet is responsible for the excess emission rather than the environment. Nevertheless, this asymmetry needs to be investigated with deeper X-ray observations to improve the significance before stronger conclusions can be drawn.

\cite{Stanley2015} have noted that 11 out of 13 hybrid sources have the X-ray jet detected on the FR\,I side. On the contrary, we find that the X-ray jet is detected on the FR\,II side of both TXS\,0730+504 and PKS\,0215+015 if indeed these sources are hybrid morphology sources. Similarly, in PKS\,1036+054 if the excess X-ray counts detected to the west are due to jet emission rather than the environment, most of our hybrid sources show X-ray jet emission on the FR\,II side of the jet. Hence, our study reduces the significance of the result that X-ray jets lie preferentially on the FR\,I side.
\section{Conclusions and Future work}
\label{sec:conclusion}

Multiwavelength observations of four hybrid morphology MOJAVE blazars are presented in this paper. We detected X-ray jet emission from two of our sources,  {PKS\,0215+015} and TXS\,0730+504.
The properties of the two sources in our sample with X-ray emission are similar to that of a general sample of X-ray jetted sources.

The non-detection of jets provided upper limits in optical and IR bands, and the resulting SED does not have a concave downward shape, consequently ruling out a single synchrotron emission model.
The SSC model requires magnetic field values well below the equipartition levels and also over-predicts $\gamma$-ray flux levels.

Compared to the synchrotron and the SSC model, the IC-CMB model reproduces the broadband SED well. The IC-CMB model also faces a few challenges such as requiring relativistic velocities on kpc-scales and over-predicting the $\gamma$-ray flux in one of the sources, TXS\,0730+504. We however note that the X-ray emission is detected only on the VLBI jet side for both our sources suggesting Doppler boosting and mildly relativistic velocities. Also, a good sampling of the low-frequency synchrotron peak is required to accurately predict the high-frequency IC-CMB peak before the IC-CMB model can be ruled out.

To be able to conclusively rule out the IC-CMB model and identify various alternative models, gathering deep supplementary data using telescopes such as  ALMA with adequate resolution is required.
Future X-ray and radio polarization observations at multiple frequencies will also help in discerning the underlying emission mechanism.
 Unlike in the IC-CMB process, higher polarization fractions are expected in X-rays if the second peak is arising from synchrotron processes \citep{Meyer2015}. Radio polarization can also offer additional constraints on the parameters improving the accuracy of the fit, for example, the electron density from rotation measure (RM) studies. Multi-scale radio polarization studies can give clues about any spine-sheath jet structure relevant for shear-layer acceleration models, and the emission region size with uniform magnetic fields. 
We find that the radio powers are similar to FR\,II type radio galaxies although they are not classified as FR\,II type sources based on morphology. These two sources further strengthens the claim that the preferred emission mechanism is dependant on the radio power rather than the FR morphology type.

\section*{Acknowledgements}
We express our sincere gratitude to the anonymous referee for their comments and suggestions which led to significant improvement of the manuscript. We thank Dr. Ethan Stanley for processing the radio data and producing the final images which were part of his PhD thesis\footnote{https://docs.lib.purdue.edu/dissertations/AAI10683102/}. We also acknowledge Dr. Aneta Siemiginowska for useful suggestions and comments. This work was supported by the National Aeronautics and Space Administration (NASA) through Chandra Award Number GO0-21089X issued by the Chandra X-ray Observatory Center (CXC), which is operated by the Smithsonian Astrophysical Observatory (SAO) for and on behalf of NASA under contract NAS8-03060. Support for HLM was provided in part by NASA through SAO contract SV3-73016 to MIT for support of the CXC.
SB and CO acknowledge support from the Natural Sciences and Engineering Research Council (NSERC) of Canada.
Support for program number HST-GO-15995.005 was provided by NASA through a grant from the Space Telescope Science Institute, which is operated by the Association of Universities for Research in Astronomy, Incorporated, under NASA contract NAS5-26555.
Support for the MOJAVE program includes NASA-Fermi grants 80NSSC19K1579, NNX15AU76G and NNX12A087G.
The National Radio Astronomy Observatory is a facility of the National Science Foundation operated under cooperative agreement by Associated Universities, Inc.

\bibliographystyle{aasjournal}
\end{document}